\documentclass[aip,amsmath,amssymb,reprint]{revtex4-1}

\usepackage{graphicx}
\usepackage{dcolumn}
\usepackage{bm}
\usepackage{color}
\usepackage[mathlines]{lineno}

\usepackage[utf8]{inputenc}
\usepackage[T1]{fontenc}
\usepackage{mathptmx}

\begin{document}

\preprint{AIP/123-QED}

\title[Propensity polarization]{A model for the competition between political mono-polarization and bi-polarization}

\author{Nicolas Saintier}
\email{nsaintie@dm.uba.ar}
\affiliation{Departamento de Matem\'atica and IMAS, UBA-CONICET, Facultad de Ciencias Exactas y Naturales, Universidad de Buenos Aires (1428) Pabell\'on I - Ciudad Universitaria - Buenos Aires - Argentina}

\author{Juan Pablo Pinasco}
\email{jpinasco@dm.uba.ar}
\affiliation{Departamento de Matem\'atica and IMAS, UBA-CONICET, Facultad de Ciencias Exactas y Naturales, Universidad de Buenos Aires (1428) Pabell\'on I - Ciudad Universitaria - Buenos Aires - Argentina}

\author{Federico Vazquez}
\email{fede.vazmin@gmail.com}
\affiliation{Instituto de C\'{a}lculo, FCEN, Universidad de Buenos Aires and CONICET, Buenos Aires, Argentina}
\homepage{https://fedevazmin.wordpress.com}

\date{\today}

\begin{abstract}
We investigate the phenomena of political bi-polarization in a population of interacting agents by means of a generalized version of the model introduced in \cite{Vazquez-2020} for the dynamics of voting intention.  Each agent has a propensity $p$ in $[0,1]$ to vote for one of two political candidates.  In an iteration step, two agents $i$ and $j$ with respective propensities $p_i$ and $p_j$ interact, and then $p_i$ either increases by an amount $h>0$ with a probability that is a nonlinear function of $p_i$ and $p_j$ or decreases by $h$ with the complementary probability.  We study the behavior of the system under variations of a parameter $q \ge 0$ that measures the nonlinearity of the propensity update rule.  We focus on the stability properties of the two distinct stationary states: mono-polarization in which all agents share the same extreme propensity ($0$ or $1$), and bi-polarization where the population is divided into two groups with opposite and extreme propensities.  We find that the bi-polarized state is stable for $q<q_c$, while the mono-polarized state is stable for $q>q_c$, where $q_c$ is a transition value that decreases as $h$ decreases.  We develop a rate equation approach whose stability analysis reveals that $q_c$ vanishes when $h$ becomes infinitesimally small.  This result is supported by the analysis of a transport equation derived in the continuum $h \to 0$ limit.  We also show by Monte Carlo simulations that the mean time $\tau$ to reach mono-polarization in a system of size $N$ scales as $\tau \sim N^{\alpha}$ at $q_c$ , where $\alpha(h)$ is a non-universal exponent. 
\end{abstract}

\maketitle

\begin{quotation}
Determining the conditions under which the opinions of a group of individuals become polarized is one of the open questions in modern sociology.  Many studies have considered different social mechanisms that lead to the emergence of opinion polarization in an interacting population, but the robustness and persistence of the polarized state were less investigated.  The present paper studies a model that incorporates a mechanism of self-reinforcement by which individuals have a tendency to strengthen their pre-existing beliefs on a given topic as they interact with others.  This leads to the emergence of extreme viewpoints in the population that can give rise to mono-polarization where everybody shares the same extreme opinion, or to bi-polarization where the population splits into two groups of opposite extreme opinions that coexist.  We give an insight into the conditions to obtain a stable bi-polarized state in terms of the specific mathematical form of the interactions between individuals.
\end{quotation}

\section{Introduction}
\label{intro}

The phenomenon of polarization of opinions about worldwide issues such as climate change, marijuana legalization, abortion and Brexit among many others, is gaining attention in recent years.  Despite the large amount of work by researchers in disciplines as diverse as social psychology, sociology, economics, political science, computer science, cognitive science and more recently physics, the conditions under which polarization arises in a society are still under vivid debate \cite{Mason-2007,Flache-2019}.  Homophily, persuasion, self-reinforcement or a combination of them have been identified in small scale experiments, tested by agent-based simulations, as different mechanisms of social influence that could give rise to polarization \cite{Mas-2013}.  When acting over a large population of individuals, these mechanisms might induce the formation of two groups with opposite viewpoints on a given topic, for instance in favor or against abortion. A distinctive feature that defines a bi-polarized population is the large agreement within the members of a group and the large disagreement between individuals of different groups.  More recently, online social media has been pointed out as to foster polarization due to the existence of echo chambers \cite{Prasetya-2019,Nguyen-2018} formed by like-minded individuals that interact between them and distrust people outside their group, or the existence of a related online phenomenon called epistemic bubbles \cite{Nguyen-2018}, where insiders do not even listen to people of other groups. 

In the last years many models have been introduced and studied in the physics literature in an attempt to explain the phenomenon of opinion polarization.  These models incorporate some of the social mechanisms described above, 
like homophily which favors interactions between like-minded individuals and may lead to the fragmentation of the interaction network in same-opinion clusters \cite{Holme-2006,Kimura-2008,Nardini-2008,Vazquez-2007-3,Vazquez-2008-1,Demirel-2014}, the theory of persuasive arguments by which two interacting individuals with the same opinion orientation reinforce their initial positions and they become more extreme \cite{LaRocca-2014,Balenzuela-2015,Velasquez-2018}, and self-reinforcement where individuals tend to intensify their opinions towards and already favored side \cite{Vazquez-2020}.  Other mechanism that generates extreme opinions is the so called bias assimilation in social psychology \cite{Lord-1979}, by which individuals exhibit a bias towards their current position when they receive new information, developing more extreme viewpoints \cite{Dandekar-2013}.  They authors in \cite{Dandekar-2013} study an extension of the DeGroot's model \cite{Degrrot-1974} and show that homophily alone is not able to induce polarization, while the addition of bias assimilation may drive the population to a polarized state when the bias is large enough.  More recently, the authors in \cite{Bhat-2020} studied an extension of the voter model where, besides adopting the opinion of a neighbor, voters are influenced by two external and opposing news sources that generate a bi-polarized population.  The authors in \cite{Schweighofer-2019} introduced a weighted balance theory for opinion formation that is able to explain not only the emergence of mono-polarization but also the phenomena of hyperpolarization that combines opinion extremeness and the correlation between issues. 

The issue about the stability of the polarized state has recently started to be investigated in some of these models.  For instance in \cite{Xia-2019} the authors study
the stability of the stationary states of the model introduced in \cite{Dandekar-2013} and found that polarization is stable for strong assimilation bias, while for intermediate bias the stability depends on the network topology.  In \cite{LaRocca-2014,Velasquez-2018} the authors show that persuasion and compromise can lead to bi-polarization, but it is unstable for all parameter values.  More recently, the work in \cite{Vazquez-2020} studies a model for the dynamics of propensities to choose between two possible political candidates in an election, which are initially
distributed uniformly in the interval $[0,1]$.  In a pairwise interaction a random agent either increases or decreases its propensity by an amount $h$ with probabilities that depend on the propensities of the two interacting agents. The propensity update includes a mechanism of self-reinforcement of one's propensity -- akin to biased assimilation -- regulated by a weight $\omega$, which tend to gradually lead agents to extreme propensity values $0$ or $1$.  Depending on the initial condition, this effect can either lead the system to a bi-polarized state characterized by two groups with extreme and opposite propensities, or to a mono-polarized state that consists of a consensus in one of the two extreme propensities.  It was shown in \cite{Vazquez-2020} by means of a linear stability analysis that the bi-polarized state is unstable (a saddle fixed point) in the entire range of parameters' values ($h>0$ and $0 < \omega \le 1$), while the mono-polarized state is always locally stable.

In this article we address the question of the robustness of bi-polarization under nonlinear modifications of the interaction rule of the propensity model studied in \cite{Vazquez-2020}.  We introduce a control parameter $q \ge 0$ that tunes the probability to increase the propensity in a pairwise interaction, which is a generalization of the linear interaction rule ($q=1$) investigated in \cite{Vazquez-2020}.  We study the dynamics of the system by means of a rate equation approach complemented with Monte Carlo simulations, and analyze the stability properties of the bi-polarized stationary state as $q$ is varied.  In particular, we ask the question on whether it is possible to obtain a stable bi-polarization.  We found that for values of $q$ smaller than a threshold value $q_c$ the bi-polarized state is stable, while it looses stability for $q>q_c$, where the value of $q_c$ depends on $h$ and $\omega$.  Simulations show that at the transition point $q_c$ the time to reach one of the two absorbing states (extreme mono-polarization) grows as a power law of $N$ with a non-universal exponent.

The outline of the paper is as follows.  We define the model and introduce the nonlinear update rule in Section~\ref{model}.  In Section~\ref{rate} we write down the rate equations for the evolution of the system for any $h$ and $\omega$.  In section~\ref{special} we study some special cases that can be handled analytically and shed some light on the effects of the modeling parameters.  In Section~\ref{S2} we investigate the simplest non-trivial case $h=1/2$.  We obtain an analytical value of the transition point $q_c$ by studying the stability properties of the fixed points and develop a potential approximation that explains the scaling of the mean consensus time with $N$ at $q_c$.  Section~\ref{MC} shows Monte Carlo (MC) simulation results.  In Section~\ref{continuous} we derive a transport equation valid in the continuum $h \to 0$ limit that gives an insight into the limiting behavior of $q_c$ as $h$ decreases.  Finally, in Section~\ref{summary} we present a summary and a short discussion of the results.

\section{Description of the model}
\label{model}

We consider an extension of the model proposed in \cite{Vazquez-2020} where each individual of a population of $N$ agents develops an inclination or propensity towards one of two possible political candidates, namely, $A$ and $B$, which is subject to change by interacting with other agents.  The propensity is represented by a real number $p$ in the interval $[0,1]$ and denotes the inclination to choose the alternative $A$, such that a value of $p$ close to $0$ ($1$) implies that the agent is very likely to vote for $B$ ($A$).  At the initial time $t=0$ each agent is assigned a random propensity in $[0,1]$, thus the propensity distribution in the population is nearly uniform.  Then, agents are allowed to change their propensities by interacting with random partners.  At each time step $\Delta t=1/N$, two agents $i$ and $j$ with respective propensities $p_i(t)$ and $p_j(t)$ at time $t$ are chosen at random. Then agent $i$ updates its propensity according to the following rule:
\begin{eqnarray} 
  p_i(t+1/N) = 
  \begin{cases}
    p_i(t)+h & \mbox{with probability $\mathbf P^+(p_i,p_j)$}, \\
    p_i(t)-h & \mbox{with probability $1-\mathbf P^+(p_i,p_j)$}, \nonumber
  \end{cases}
  \label{pi}
\end{eqnarray}
where 
\begin{equation}
  \mathbf P^+(p_i,p_j) = \frac{p_{i,j}^q}{p_{i,j}^q+(1-p_{i,j})^q}, ~~ \mbox{and} ~~ p_{i,j}=\omega p_i+(1-\omega) p_j. 
  \label{Pij}
\end{equation}
That is, the propensity $p_i$ of agent $i$ either increases by a fixed step length $h$ ($0 < h < 1$) with a probability $\mathbf P^+(p_i,p_j)$ that is a nonlinear function of the weighted average $p_{i,j}$ of both propensities $p_i$ and $p_j$, or decreases by $h$ with the complementary probability $1-\mathbf P^+$.  If $p_i$ gets larger (smaller) than $1$ ($0$) its value is reset to $1$ ($0$), thus propensities are always contained in the interval $[0,1]$.  This time step is repeated ad infinitum. The parameter $\omega$ ($0 \le \omega \le 1$) is the weight that an agent gives to its own propensity in a pairwise interaction, and is the same for all agents. The nonlinearity of $\mathbf P^+$ is controlled by the parameter $q \ge 0$, so that the linear case $q=1$ corresponds to the dynamics studied in \cite{Vazquez-2020}.

\begin{figure}[t]	
  \includegraphics[width=\columnwidth]{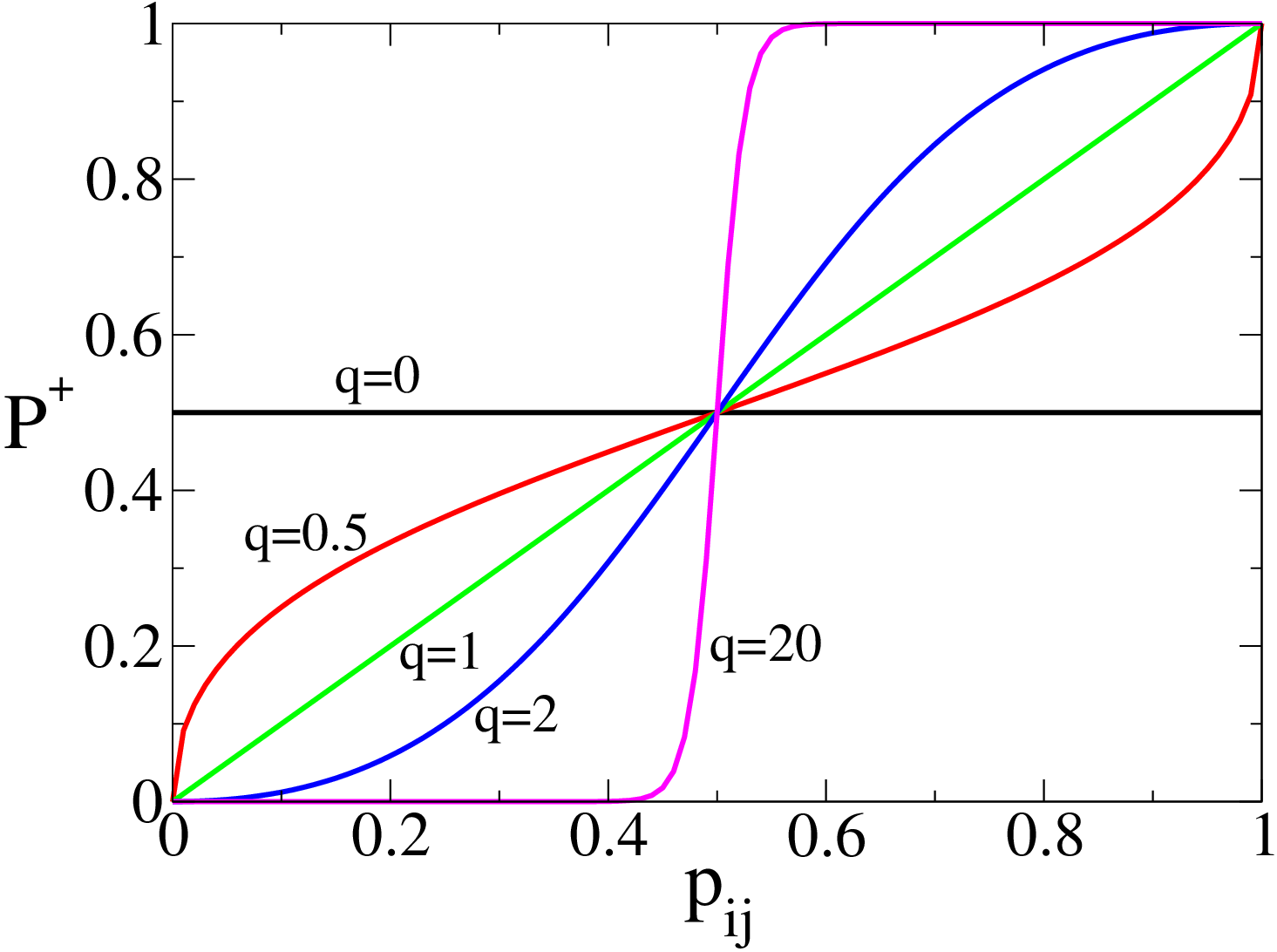}
  \caption{Probability $\mathbf P^+$ that an agent increases its propensity $p_i$ by $h$ when interacts with an agent with propensity $p_j$ as a function of their weighted average propensity $p_{i,j}$ from Eq.~(\ref{Pij}), for the values of $q$ indicated in the plot.}
  \label{fig:ProbaTran_q}
\end{figure}

To obtain an intuition about how $q$ affects the probability to increase the propensity in an interaction, in Fig.~\ref{fig:ProbaTran_q} we plot $\mathbf P^+$ as a function of $p_{i,j}$ for different values of $q$.  Notice that, for all $q \ge 0$, $\mathbf P^+=0$ ($\mathbf P^+=1$) when both interacting agents have the same extreme propensity $p_i=p_j=0$ ($p_i=p_j=1$).  In the limiting case $q=0$ is $\mathbf P^+= 1/2$ for any $p_i$ and $p_j$ except when $p_{i,j}=0$ or $p_{i,j}=1$ as mentioned above.  Then, agents perform independent symmetric random walks in $(0,1)$ with ``elastic'' boundaries at $0$ and $1$, and thus the population's propensity distribution tends to remain uniform.  In the opposite limit $q \to \infty$, $\mathbf P^+$ tends to the step function $\Theta(p_{i,j}-1/2)$, i e., $\mathbf P^+=0$ ($\mathbf P^+=1$) for $p_{i,j}<1/2$ ($p_{i,j}>1/2$), and $\mathbf P^+(1/2)=1/2$.  As as consequence, a pair of interacting agents with propensities smaller than $1/2$, and so with a weighted average propensity $p_{i,j} <1/2$, would most likely decrease their propensities, approaching the lowest value $p=0$.  Equivalently, their propensities would approach $p=1$ if both agents have propensities larger than $1/2$.  This generates a mechanism of reinforcement in which individuals with the same opinion orientation become more extreme as they interact, which is known in the literature to induce bi-polarization \cite{Mas-2013,LaRocca-2014,Velasquez-2018}.

\section{Rate equations}
\label{rate}

Initially, propensities take a uniform random real value in the interval $[0,1]$.  As the system evolves under the rules of section~\ref{model}, after a short time all agents adopt propensities that are multiples of $h$, i e., $p=0,h,2h,..1$, due to boundary effect that forces propensities to remain in $[0,1]$.  For the sake of simplicity in our analysis we consider a step $h$ such that $S \equiv 1/h$ is an integer number, and thus propensities adopt discrete values $p=kh$, with $k=0,..,S$.  Then, for times $t \gtrsim 1$ the propensity distribution can be written as $f(p,t)=\sum_{k=0}^S n_k(t) \, \delta(p-kh)$, where $n_k(t)$ is the fraction of agents with propensity $kh$ (in state $k$ from now on) at time $t$, and where $\sum_{k=0}^S n_k(t)=1$ for all times $t \ge 0$.  Then, the time evolution of the fractions $n_k$ can be described by the following system of coupled rate equations:
\begin{subequations}
  \begin{alignat}{3}
    \label{dndt0}
    \frac{dn_0}{dt} &= n_1\sum_{k'=0}^S n_j (1-P_{1,k'}) - n_0 \sum_{k'=0}^S n_{k'} P_{0,k'}, \\
    \label{dndtk}
    \frac{dn_k}{dt} &= n_{k-1} \sum_{k'=0}^S n_{k'} P_{k-1,k'} + n_{k+1} \sum_{k'=0}^S n_{k'} (1-P_{k+1,k'}) - n_k \\ & \mbox{for $1 \le k \le S-1$}, \nonumber \\
    \label{dndtS}
    \frac{dn_S}{dt} &=  n_{S-1} \sum_{k'=0}^S n_{k'} P_{S-1,k'} - n_S \sum_{k'=0}^S n_{k'} (1-P_{S,k'}), 
  \end{alignat}
  \label{dndt}
\end{subequations}
where 
\begin{equation}
  P_{k,k'} = \frac{p_{k,k'}^q}{p_{k,k'}^q+(1-p_{k,k'})^q}, ~~ \mbox{and} ~~ p_{k,k'}=\omega k h+(1-\omega) k' h. 
  \label{Pkk'}
\end{equation}
Here the short notation $P_{k,k'} \equiv  P_{k,k'}(k \to k+1) = P_{k,k'}^+$ denotes the transition probability of an agent (a particle from now on) from state $k$ to state $k+1$ or, equivalently, from propensity $kh$ to $(k+1)h$, when it  interacts with another particle with state $k'$ (or propensity $k'h$).  We have also dropped the $+$ sign in $P_{k,k'}^+$ to simplify notation.  Analogously, the  transition $k \to k-1$ happens with the complementary probability $1-P_{k,k'} \equiv P_{k,k'}(k \to k-1)$.  The first and second gain terms in Eq.~(\ref{dndtk}) correspond to the transitions of particles from state $k-1$ to state $k$ and from state $k+1$ to state $k$, respectively.  For instance, the transition event $k-1 \to k$ happens when a particle in state $k-1$ is chosen at random (with probability $n_{k-1}$), then interacts with a particle in state $j$ (with probability $n_j$), and jumps to state $k$ with probability $P_{k-1,j}$.  Adding over all possible values of $j=0,..,S$ yields the first term.  Similarly, the transition $k+1 \to k$ happens when a $k+1$ particle interacts with a $j$ particle and then jumps to state $k$ with probability $1-P_{k+1,j}$.  The third (loss) term represents $k \to k-1$ and $k \to k+1$ transitions, which happen with probability $1$ after interacting with any other particle. 

We note that the mono-polarized states $(n_0,..,n_S)=(1,0,...,0)$ and $(n_0,..,n_S)=(0,...,0,1)$ are fixed points of the rate Eqs.~(\ref{dndt}), and represent a consensus state in the extreme propensity $p=0$ and $p=1$, respectively.  As we shall see later, there is a third non-trivial fixed point that corresponds to a bi-polarized state in which most particles adopt extreme propensity values. 

The aim of this article is to study by analytical and numerical means the stability of these states, and how that depends on the values of the parameters $q$, $\omega$ and $N$.  This study makes sense from the viewpoint of the  statistical physics of critical phenomena since the mono-polarized state is stable for the linear case $q=1$ as we know from the work in \cite{Vazquez-2020}, but it is unstable for small enough values of $q$, as we shall see in detail in section~\ref{stability} for the $S=2$ case.  Therefore, as $q$ decreases from $1$ we expect to find a threshold value $q_c$ below which the mono-polarized state becomes unstable, leading to a transition from a phase where particles adopt the same extreme propensity at the steady state (mono-polarization) to a phase with a stable coexistence of opposite and extreme propensities (bi-polarization).
  
Stochastic fluctuations in a finite system eventually lead to one of the absorbing mono-polarized states, and thus we also aim to study the time to reach mono-polarization or extreme consensus.  We perform an analytical exploration of the stability and consensus times in the limiting cases $\omega=0$, $\omega=1$ and $q=0$ for all $S$ in Section~\ref{special}, as well as in the simplest non-trivial case $S=2$ in Section~\ref{S2}, which is amenable to theoretical analysis and contains most of the observed phenomenology of the system.  Consensus times for $S \ge 3$ are analyzed by MC simulations in Section~\ref{MC}, and the continuous case $S \to \infty$ is investigated analytically in Section~\ref{continuous}.  Notice that the stability analysis of the system and the estimation of consensus times are not trivial tasks due to the nonlinearity of the interaction rules.

\section{Special cases}
\label{special}

In this section we study some particularly simple cases, namely $\omega=0$, $\omega=1$ and $q=0$, where the rate Eqs.~(\ref{dndt}) are reduced to much simpler forms.  We focus on the non-trivial fixed points, i e., $(n_1^*,..,n_S^*) \neq (1,0,..,0)$ and $(n_1^*,..,n_S^*) \neq (0,..,0,1)$.  At the same time, these limiting cases allow to gain some intuition into the system's dynamics before studying the full nonlinear version of the model. 

\subsection{Case $\omega=0$} 
\label{w0}

For $\omega=0$ is $P_{k,k'}=\frac{k'^q}{k'^q+(S-k')^q} \equiv P_{k'}$.  Denoting by $\langle P \rangle \equiv \sum_{k'=0}^S n_{k'} P_{k'}$ the mean value of $P_{k'}$, the rate Eqs.~(\ref{dndt}) are reduced to 
\begin{eqnarray*}
  \frac{dn_0}{dt} &=& (1-\langle P\rangle)n_1 - \langle P\rangle n_0, \\
  \frac{dn_k}{dt} &=& \langle P\rangle n_{k-1} + (1-\langle P\rangle) n_{k+1} - n_k \qquad \mbox{for $1 \le k \le S-1$}, \\
  \frac{dn_S}{dt} &=&  \langle P\rangle n_{S-1} - (1-\langle P\rangle)n_S. 
\end{eqnarray*}
These equations describe the evolution of a system of random walkers with hopping probabilities $\langle P \rangle$ and $1-\langle P \rangle$ to the right and left neighboring sites, respectively, of a one-dimensional chain $[0,h,2h,..,1]$.  The stationary state of the system is given by the solution of the set of equations
\begin{eqnarray*}
  0 &=& (1-a)n_1^* - a n_0^*, \\
  0 &=& a n_{k-1}^* + (1-a) n_{k+1}^* - n_k^* \qquad \mbox{for $1 \le k \le S-1$}, \\
  0 &=&  a n_{S-1}^* - (1-a) n_S^*,
\end{eqnarray*}
where $n_k^*$ and $a$ denote the stationary value of $n_k$ and $\langle P \rangle$, respectively. Solving these equations recursively from $n_0^*$ and also from $n_S^*$ we obtain 
\begin{eqnarray*}
  n_k^* &=& \left( \frac{a}{1-a} \right)^k n_0^* \qquad \text{and} \\
  n_k^* &=& \left( \frac{1-a}{a} \right)^{S-k} n_S^*,
\end{eqnarray*}
respectively.  Since $n_k^* \in [0,1]$, then for $S \gg 1$ we must have $\frac{a}{1-a} \le 1$ and $\frac{1-a}{a} \le 1$, and thus $a=1/2$.  Then, all $n_k^*$ are equal and from the normalization condition we obtain $n_k^*=1/(S+1)$ for all $k=0,..,S$.  Therefore, the non-trivial stationary state corresponds to a uniform distribution of propensities.

\subsection{Case $\omega=1$} 

For $\omega=1$ is $P_{k,k'}=\frac{k^q}{k^q+(S-k)^q} \equiv P_{k}$.  Note that, in particular, $P_0=0$ and $P_S=1$.  Then, the rate Eqs.~(\ref{dndt}) become
\begin{subequations}
  \begin{alignat}{3}
    \frac{dn_0}{dt} &= (1-P_1^+) n_1, \\
    \frac{dn_k}{dt} &= P_{k-1} n_{k-1}  + (1-P_{k+1}) n_{k+1} - n_k \qquad  \mbox{for $1 \le k \le S-1$},  \\
    \frac{dn_S}{dt} &= P_{S-1} n_{S-1}. 
  \end{alignat}
  \label{dndt-w1}
\end{subequations}
These equations represent a system of random walkers in the chain $[0,h,2h,..,S]$ with right and left hopping probabilities $P_k$ and $1-P_k$, respectively.  To find the stationary fractions $n_k^*$ from Eqs.~(\ref{dndt-w1}) we set the time derivatives to $0$ and solve them recursively to obtain $n_k^*=0$ for $k=1,2,..,S-1$, and with no restrictions on $n_0^*$ and $n_S^*$, except $n_0^*+n_S^*=1$.  Therefore, at the stationary state each agent has either propensity $0$ or $1$ and the fractions of particles in each propensity state $n_0^*$ and $n_S^*$ depend on the initial condition.  For instance for $S=2$ we have $P_1=1/2$ and thus the system of Eqs.~(\ref{dndt-w1}) become 
\begin{eqnarray*}
  \frac{dn_0}{dt} = \frac{1}{2} n_1, \qquad \frac{dn_1}{dt} = - n_1,  \qquad
  \frac{dn_2}{dt} = \frac{1}{2} n_1,
\end{eqnarray*}
whose solution is
\begin{eqnarray*}
  n_0(t) &=& n_0(0)+\frac{1}{2} n_1(0)(1-e^{-t}), \\
  n_1(t) &=& n_1(0) e^{-t}, ~~ \mbox{and} \\
  n_2(t) &=& n_2(0)+\frac{1}{2} n_1(0)(1-e^{-t}). 
\end{eqnarray*}
Therefore, in the long time limit we obtain $n_0^* = n_0(0)+ \frac{1}{2} n_1(0)$ and $n_2^* = n_2(0)+ \frac{1}{2} n_1(0)$. \\

In the rest of the article we assume that $0 < \omega < 1$ ($\omega \ne 0$ and $\omega \ne 1$) unless otherwise stated.

\subsection{Case $q=0$}

For $q=0$ is $P_{0,0}=0$, $P_{S,S}=1$ and $P_{k,k'}=1/2$ for any $(k,k') \ne (0,0)$ and $(S,S)$.  Then, the rate Eqs.~(\ref{dndt}) are
\begin{subequations}
  \begin{alignat}{5}
    \frac{dn_0}{dt} &= \frac12 n_1 - \frac12 n_0(1-n_0), \\ 
    \frac{dn_1}{dt} &= \frac12 n_0(1-n_0) + \frac12 n_2 - n_1, \\ 
    \frac{dn_k}{dt} &= \frac12 n_{k-1} + \frac12 n_{k+1} - n_k \qquad  \mbox{$2 \le k \le S-2$},\\
    \frac{dn_{S-1}}{dt} &= \frac12 n_{S-2} + \frac12 n_S(1-n_S) - n_{S-1}, \\ 
    \frac{dn_S}{dt} &= \frac12 n_{S-1} - \frac12 n_S(1-n_S). 
  \end{alignat}
  \label{dndt-q1}
\end{subequations}
Notice that the equations for $k=2,..,S-2$ correspond to that of a symmetric random walker that jumps to the right and left with probabilities equal to $1/2$, whereas the equations for $k=0,1,S-1$ and $S$ are slightly different because they reflect the bouncing effect at the boundaries $p=0$ and $p=1$.  We thus expect to find a nearly uniform distribution of propensities at the stationary state, as we show below.  Indeed, setting the left-hand side of Eqs.~(\ref{dndt-q1}) to $0$ and solving for the stationary values $n_k^*$ we find
\begin{eqnarray}
  n_1^*=n_2^*=...=n_{S-1}^*=n_0^*(1-n_0^*)=n_S^*(1-n_S^*).
  \label{n0nS}
\end{eqnarray}
Then, the normalization condition $\sum_{k=0}^S n_k=1$ becomes
\begin{eqnarray}
  n_S^*+Sn_0^*(1-n_0^*)+n_0^{*^2} = n_0^* + S n_S^*(1-n_S^*)+n_S^{*^2} = 1.
  \label{normal}
\end{eqnarray}
Besides, from Eq.~(\ref{n0nS}) we arrive to the simple relation
\begin{eqnarray*}
  (n_0^*-n_S^*)(n_0^*+n_S^*-1)=0.
\end{eqnarray*}
Then, if $n_0^*+n_S^*=1$ we obtain that $n_1^*=...=n_{S-1}^*=0$ and from Eq.~(\ref{n0nS}) that $n_0^* \, n_S^*=0$, and thus $n_0^*=0$ ($n_S^*=1$) or $n_S^*=0$ ($n_0^*=1$), which represent the two mono-polarized state solutions.  If $n_0^*=n_S^*$, then Eq.~(\ref{normal}) leads to the following quadratic equation for $n_0^*$:
\begin{eqnarray}
n_0^* + Sn_0^*(1-n_0^*)+n_0^{*^2} = 1,
\end{eqnarray}
with solution
\begin{eqnarray}
  n_0^* = n_S^* = \frac{S+1-\sqrt{(S-1)^2+4}}{2(S-1)},
  \label{n0nS*}
\end{eqnarray}
and $n_1^*=n_2^*=...=n_{S-1}^*=n_0^*(1-n_0^*)<n_0^*$.  For $S \gg 1$ we can check from Eq.~(\ref{n0nS*}) that $n_0^*=n_S^* \gtrsim 1/S=h$, and thus $n_k \lesssim h$ for $k=1,..,S-1$.  That is, the propensity distribution is uniform in the interval $[h,1-h]$ and slightly peaked at the boundaries $p=0$ and $p=1$, and approaches the uniform distribution as $S \to \infty$.

\section{Case $S=2$}
\label{S2}

\subsection{Stationary states}

A first insight into the behavior of the model can be obtained by studying the simplest non-trivial case $S=2$ ($h=1/2$) where the propensity can take one of three possible values $p=0,1/2$ and $1$.  From Eq.~(\ref{Pkk'}), the transition probabilities $P_{k,k'}$ ($k,k' \in \{0,1,2\}$) are 
\begin{eqnarray}
 P_{0,0} &=&0, ~~~ P_{0,1}=\frac{(1-\omega)^q}{(1-\omega)^q+(1+\omega)^q}, ~~~ P_{0,2} =\frac{(1-\omega)^q}{(1-\omega)^q+\omega^q}, \nonumber \\ 
 P_{1,0} &=&\frac{\omega^q}{\omega^q+(2-\omega)^q}, ~~~  P_{1,1} =\frac12, ~~~ P_{1,2} = 1-P_{1,0}, \nonumber \\ 
  P_{2,0} &=& 1-P_{0,2}, ~~~ P_{2,1}=1-P_{0,1}, ~~~ P_{2,2}=1.  
  \label{Ps}
\end{eqnarray} 
Since $n_1=1-n_0-n_2$, it is convenient to work with the following closed systems of equations for $n_0$ and $n_2$ obtained from Eqs.~(\ref{dndt}):
\begin{subequations}  
  \begin{alignat}{2}
    \label{dndt0-2}
    \frac{dn_0}{dt} &= n_0n_1(1-P_{1,0}-P_{0,1}) + \frac12 n_1^2 + n_1n_2 P_{1,0} - n_0n_2 P_{0,2}, \\
    \label{dndt2-2}
    \frac{dn_2}{dt} &= n_0n_1 P_{1,0} + \frac12 n_1^2 + n_1n_2 (1-P_{1,0}-P_{0,1}) - n_0n_2 P_{0,2}.
  \end{alignat}
  \label{dndt-2}
\end{subequations}
The fixed points of the system are given by the solutions of 
\begin{subequations}  
  \begin{alignat}{2}
    \label{n0n2-1}
    n_0^* \, n_1^*(1-P_{1,0}-P_{0,1}) + \frac12 n_1^{*^2} + n_1^* \, n_2^* P_{1,0} - n_0^* \, n_2^* \, P_{0,2} &=& 0, \\
    \label{n0n2-2}
    n_0^* \, n_1^* \, P_{1,0} + \frac12 n_1^{*^2} + n_1^* \, n_2^* \, (1-P_{1,0}-P_{0,1}) - n_0^* \, n_2^* \, P_{0,2} &=& 0.
  \end{alignat}
  \label{n0n2}
\end{subequations}
Subtracting Eq.~(\ref{n0n2-2}) from Eq.~(\ref{n0n2-1}) gives $(1-2P_{1,0}-P_{0,1})n_1(n_0-n_2)=0$.  This relation is satisfied when either (i) $n_1^*=0$ or (ii) $1-2P_{1,0} -P_{0,1}=0$ or (iii) $n_0^*=n_2^*$.  Case (i) gives the two trivial fixed points corresponding the mono-polarized states $(n_0^*,n_2^*)=(1,0)$ or $(0,1)$, as we can check by setting $n_1^*=0$ ($n_0^*+n_2^*=1$) into Eq.~(\ref{n0n2-1}), which leads to solutions $n_0^*=0$ or $n_0^*=1$.  As we shall see in the next section, condition (ii) is fulfilled at
the transition points $q_c(\omega)$ in the $(\omega,q)$ plane for which the stability of the mono-polarized states changes.  Finally, case (iii) corresponds to the non-trivial fixed point that describes a bi-polarized state, as we show below.  If $n_0^*=n_2^*$ then $n_1^*=1-2n_0^*$ and thus we can see from Eq.~(\ref{n0n2-1}) that $n_0^*$ solves the quadratic equation  
$(2P_{0,1}-P_{0,2})x^2 - (1+P_{0,1}) x + 1/2=0$.  Notice from relations in Eqs.~(\ref{Ps}) that $2P_{0,1}-P_{0,2}= 0$ only when $q=1$, in which case $n_0^*=n_2^*=1/(3-\omega)$.  When $q\neq 1$, the  roots of the quadratic equation are 
\begin{equation*}
x^\pm = \frac{1+P_{0,1} \pm \sqrt{(1-P_{0,1})^2 + 2P_{0,2}}}{2(2P_{0,1}-P_{0,2})}. 
\end{equation*}
We observed numerically that $x^+>1$ and $x^-\in (0,1)$ for any $\omega\in [0,1]$ and $q\ge 0$.  Thus, the solution with physical meaning is   
\begin{equation}
  n_0^*=n_2^* = \frac{1+P_{0,1} - \sqrt{(1-P_{0,1})^2 + 2P_{0,2}}}{2(2P_{0,1}-P_{0,2})}.
  \label{n0-n2-sol}
\end{equation}
Using expressions for $P_{0,1}$ and $P_{0,2}$ from Eqs.~(\ref{Ps}) we have that   in the $\omega \to 0$ limit is $P_{0,1}=\frac12 - \frac{q\omega}{2}+O(\omega^3)$ and  $P_{0,2}=1-\omega^q+O(\omega^3)$, thus one can check from Eq.~(\ref{n0-n2-sol}) that $(n_0^*,n_1^*,n_2^*) \to (1/3,1/3,1/3)$ as $\omega \to 0$, corresponding to the uniform distribution case discussed in Section~\ref{w0}.  Now, in the opposite limit $\omega \to 1$ we have $P_{0,1} \ll 1$ and $P_{0,2} \ll 1$ (see Appendix~\ref{1-w-1} for Taylor expansion details), and so $(n_0^*,n_1^*,n_2^*) \to (1/2,0,1/2)$.  This represents a totally bi-polarized state that is symmetric respect to $p=1/2$, where all agents adopt one of the two possible extreme propensities $p=0$ or $p=1$ in equal proportions.  For intermediate values of $\omega>0$ the fixed point from Eq.~(\ref{n0-n2-sol}) corresponds to a symmetric bi-polarized state where most agents adopt extreme propensities ($n_0^*=n_2^* > n_1^*$).  For instance, for $\omega=1/2$ is
\begin{equation}
  n_0^*(\omega=1/2) = \frac{2+3^q \pm \sqrt{1+2 \times 3^q(1+3^q)}}{3-3^q}.
\end{equation}
The fraction of agents at the extreme propensities $n_0^*=n_2^*$ is larger than $n_1^*$ for all $q \ge 0$ and increases very slowly with $q$.  For instance $(n_0^*,n_1^*,n_2^*)|_{q=0}^{\omega=0.5} \simeq (0.382,0.236,0.382)$, 
$(n_0^*,n_1^*,n_2^*)|_{q=1}^{\omega=0.5} \simeq (0.4,0.2,0.4)$ and
$(n_0^*,n_1^*,n_2^*)|_{q \to \infty}^{\omega=0.5} \to (0.414,0.172,0.414)$.

Even though we showed above that for $S=2$ there are three fixed points, two  corresponding to mono-polarization and one to bi-polarization, we shall see in the next section that only one is stable for a given set of $\omega$ and $q$ values. Determining the stable fixed point is important because it represents the true stationary state in a finite system.  This is so because finite-size fluctuations play the role of small perturbations of the trajectories of $n_k(t)$ that eventually take the system away from an unstable configuration, corresponding to an unstable fixed point, and leads it towards a stable stationary configuration.

\subsection{Stability analysis}
\label{stability}

\begin{figure}[t]	
  \includegraphics[width=\columnwidth]{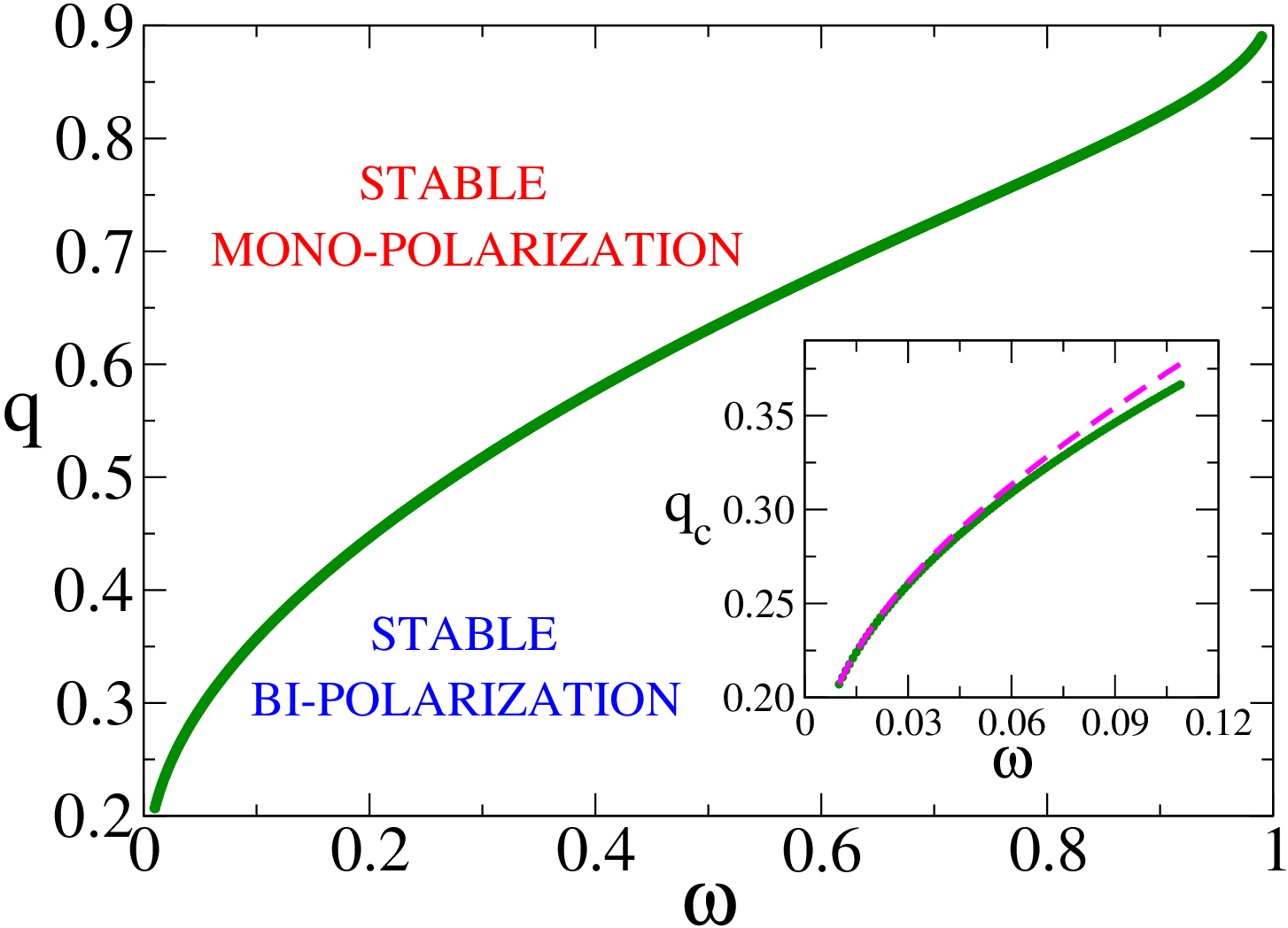}
  \caption{Phase diagram in the $\omega-q$ space for $S=2$ obtained from the stability analysis.  The transition curve $q_c(\omega)$ separates the bi-polarization phase ($q>q_c$) from the mono-polarization phase ($q<q_c$).  The inset shows the behavior of $q_c$ for small values of $\omega$.  The dashed line is the analytical approximation $q_c \simeq \ln(3)/\ln(2/\omega)$.}
  \label{qc-w}
\end{figure}

To obtain a complete stability picture of the fixed points described above is enough to study the stability of the mono-polarized state $(n_0^*,n_1^*,n_2^*)=(1,0,0)$, due to the fact that for a given set $(\omega,q)$ the bi-polarized state is stable when the mono-polarized is unstable and vice-versa.  For that, we linearize the system of Eqs.~(\ref{dndt-2}) around $(n_0^*,n_2^*)=(1,0)$:
\begin{equation}
  \frac{d {\bf n}}{dt} = {\bf A n},
\end{equation}
where
\begin{eqnarray}
  {\bf A} = \begin{pmatrix}
    P_{1,0}+P_{0,1}-1 & 1-P_{1,0}-P_{0,1}+P_{0,2} \\ 
    P_{1,0} & -P_{1,0}-P_{0,2}
  \end{pmatrix},
\end{eqnarray}
and ${\bf n}=(n_0,n_2)$.  The eigenvalues of the matrix ${\bf A}$ are real numbers given by
\begin{equation}
  \lambda_{\pm} = \frac{Tr({\bf A}) \pm \sqrt{Tr^2({\bf A})-4 \, det({\bf A})}}{2},
\end{equation}
where $Tr({\bf A})=P_{0,1}-P_{0,2}-1 = -P_{2,1}-P_{0,2}<0$ is the trace of ${\bf A}$ and $det({\bf A})=P_{0,2}(1-2P_{1,0}-P_{0,1})$ is its determinant.  Using that $\lambda_- < 0$ and the relation $\lambda_+ \lambda_-=det({\bf A})$ we have that $\lambda_+<0$ when $det({\bf A})>0$ and, therefore, both eigenvalues are negative and thus the fixed point $(n_0^*,n_2^*)$ is stable.  Analogously, $\lambda_+>0$ when $det({\bf A})<0$ and thus $(n_0^*,n_2^*)$ is unstable (a saddle fixed point).  Then, the stability of the mono-polarized state changes when $det({\bf A})$ changes sign or, equivalently, when
\begin{eqnarray}
  \label{f-w-q}
  f(\omega,q) & \equiv & 1-2P_{1,0}-P_{0,1} \\ 
  & = & 1 - \frac{2 \omega^q}{\omega^q+(2-\omega)^q} -  \frac{(1-\omega)^q}{(1-\omega)^q+(1+\omega)^q} = 0. \nonumber
\end{eqnarray}
Here $f(\omega,q)$ is a function of $\omega$ and $q$ that increases with $q$ from the value $f(\omega,q=0)=-1/2$ up to $f(\omega,q=1)=(1-\omega)/2>0$, for any $\omega\in (0,1)$.  Thus, the equation $f(\omega,q)=0$ has a unique solution $q_c$ in $(0,1)$.  Then, the mono-polarized state $(1,0,0)$ is stable for $q>q_c$ and unstable for $q<q_c$.  In Fig.~\ref{qc-w} we show the curve $q_c(\omega)$ corresponding to the numerical solution of Eq.~(\ref{f-w-q}).

For the case $\omega=1/2$ the expression for $f(\omega,q)$ is rather simple, $f(1/2,q)=(3^q-2)/(3^q+1)$, from where we obtain
\begin{equation}
  q_c(\omega=1/2)=\frac{\ln(2)}{\ln(3)} \simeq 0.63093.
  \label{qc-0.5}
\end{equation}

Even though solving for $q_c$ in Eq.~(\ref{f-w-q}) is hard in general, we can still obtain an approximate expression for $q_c(\omega)$ in the $\omega \to 0$ limit.  Indeed, taking $\omega \ll 1$ and $q<1$ is $\omega \ll \omega^q$ 
and thus $f$ can be approximated as
$f(\omega,q)\simeq \frac12 - \frac{2 \omega^q}{\omega^q+ 2^q}$.  Setting $f(\omega,q)=0$ and solving for $q$ we find $q_c(\omega) \simeq \ln(3)/\ln(2/\omega)$, which is a good approximation when $\omega \lesssim 0.1$ (dashed line in the inset of Fig.~\ref{qc-w}) and shows that $q_c \to 0$ as $\omega \to 0$. 

Summarizing, for the $S=2$ case we show that as $q$ increases from $0$ there is a transition from a phase in which bi-polarization is stable ($q<q_c$) to a phase where mono-polarization is stable ($q>q_c$), corresponding to the stationary states in a finite system.

\subsection{Time to consensus}
\label{consensus}

An interesting magnitude to study in these systems is the mean time to reach the final absorbing configuration, i e., a mono-polarized state or extremist consensus.  The stability analysis performed in section~\ref{stability} is also useful to estimate the mean consensus time $\tau$ because it allows to relate $\tau$ with the eigenvalues of the matrix ${\bf A}$.  Starting from a state where the $p=0$ stable consensus $(1,0,0)$ is slightly perturbed ($n_0^* \lesssim 1$), the linear stability analysis shows that, at long times, the system goes back to consensus following the exponential relaxation $1-n_0(t) \sim e^{\lambda_{\mbox{\tiny max}} t}$, where $\lambda_{\mbox{\scriptsize max}}<0$ is the largest of the two eigenvalues $\lambda_{\pm} < 0$.  We are assuming here that the system is in the mono-polarized phase $q>q_c$, and thus both eigenvalues $\lambda_+$ and $\lambda_-$ are negative.  Then, we can estimate the mean consensus time in a system of $N$ particles as the time $\tau$ for which $n_0$ becomes larger than $1-1/N$ (less than one particle with state $k \ne 0$), that is, $1/N \simeq 1-n_0(\tau) \sim e^{\lambda_{\mbox{\tiny max}} \tau}$, from where we obtain the simple approximate expression
\begin{eqnarray}
  \tau \sim -\frac{\ln N}{\lambda_{\mbox{\scriptsize max}}}.
  \label{tau-lambda}
\end{eqnarray}
  
Analytical expressions for $\lambda_{\pm}$ can be obtained in two different limits (see Appendix~\ref{stability-2}):
\begin{itemize}
  
 \item $q \gg 1$ limit: we obtain in Appendix~\ref{wgg1} that for $\omega < 1/2$ is $\lambda_{\pm} \simeq -1$, while for $\omega>1/2$ is $\lambda_+ \simeq -\left( \frac{1-\omega}{\omega} \right)^q$ and $\lambda_- \simeq -1$.  Therefore, the mean consensus time should scale as 
\begin{eqnarray}
  \tau \sim 
  \begin{cases} 
    \ln N & \mbox{for $\omega<1/2$}, \\ 
    \left( \frac{\omega}{1-\omega} \right)^q \ln N & \mbox{for $\omega>1/2$}. 
  \end{cases}
  \label{tau-q-gg-1}
\end{eqnarray}

\item $q>1$ and $1-\omega \ll 1$ limit: we obtain in Appendix~\ref{1-w-1} that $\lambda_+ \simeq - q (1-\omega)^{q+1}$ and $\lambda_- \simeq -1$ and thus
\begin{equation}
  \tau \sim ~\frac{\ln N}{q(1-\omega)^{q+1}}.
  \label{tau-q-g-1}
\end{equation}
  
\end{itemize}

\subsection{Potential approach}
\label{potential}

We develop here an approach that attempts to describe the dynamics of the model in terms of a rate equation for the time evolution of a single macroscopic variable, i e., the mean value of the propensity $m \equiv \langle p \rangle = h \sum_{k=0}^S k \, n_k$ over the population.  For the sake of simplicity we focus on the particular case $\omega=1/2$, but the same approach can be done for any $\omega$.  For $\omega=1/2$ is $P_{0,1}=P_{1,0}=1/(1+3^q)$ and $P_{0,2}=1/2$, thus the rate Eqs.~(\ref{dndt-2}) reduce to 
\begin{subequations}
  \begin{alignat}{2}
    \label{dndt0-3}
    \frac{dn_0}{dt} &= (1-2P_{0,1}) n_0 n_1 + \frac{1}{2} n_1^2 + P_{0,1} n_1 n_2 - \frac{1}{2} n_0 n_2, \\
    \label{dndt2-3}
    \frac{dn_2}{dt} &= P_{0,1} n_0 n_1 + \frac{1}{2} n_1^2 + (1-2P_{0,1}) n_1 n_2 - \frac{1}{2} n_0 n_2.
  \end{alignat}
  \label{dndt-3}
\end{subequations}
Then, the mean propensity $m \equiv \frac{1}{2} n_1 + n_2 = \frac{1}{2} (1+n_2-n_0)$ evolves according to 
\begin{equation}
  \frac{dm}{dt} = (1-3P_{0,1})\, n_1 \left( m-1/2 \right).
  \label{dmdt}
\end{equation}
This equation shows that the fixed point $m=1/2$ corresponding to the symmetric polarized state is stable when $P_{0,1} > 1/3$ and unstable when $P_{0,1} < 1/3$.  The stability transition takes place when $P_{0,1}=1/3$, that is at $q_c=\ln(2)/\ln(3)$ as found by the stability analysis [Eq.~(\ref{qc-0.5})].   It is interesting to note from Eq.~(\ref{dmdt}) that for $q=q_c$ the mean propensity is conserved ($dm/dt=0$) for all times $t \ge 0$.  As there is no net drift over $m$, the evolution of $m$ in a finite system is only driven by finite-size fluctuations, which are usually described by an additional noise term that is absent in the present rate equation formalism for infinite large systems.  This behavior is reminiscent of that of the voter model \cite{Holley-1975,Clifford-1973} where the magnetization is conserved at every step of the dynamics and one of the two absorbing consensus state is reached by fluctuations \cite{Vazquez-2008-2}.  Therefore, it is expected that the consensus time for the $S=2$ case scales as in the voter model in mean-field \cite{}, i e., $\tau \sim N$, as we shall confirm by MC simulations in section~\ref{MC}.

\begin{figure}[t]
  \includegraphics[width=\columnwidth]{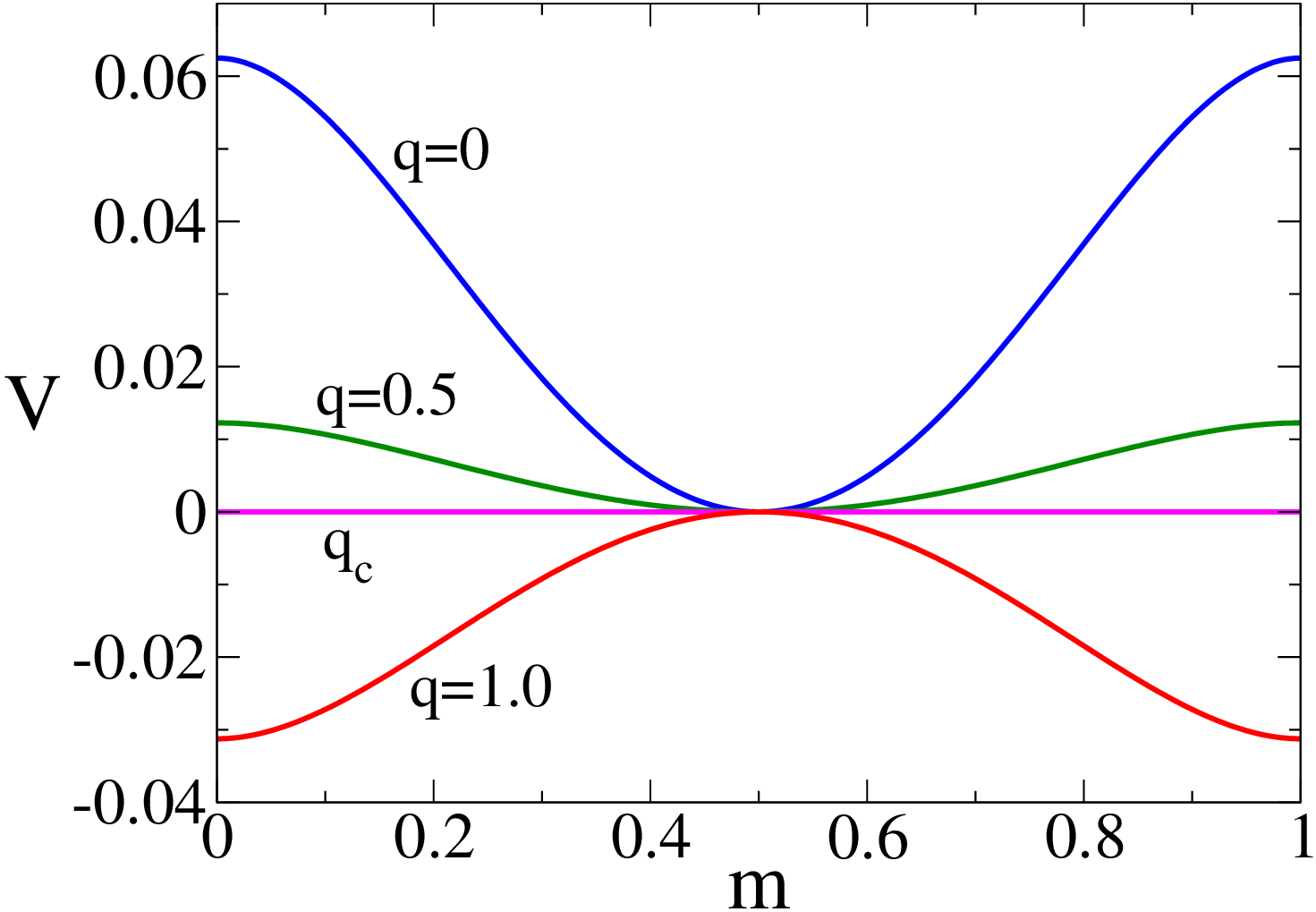}
  \caption{Ginzburg-Landau Potential $V$ vs mean propensity $m$ from Eq.~(\ref{Vm}) for $q=0, 0.5, \ln(2)/\ln(3)$ and $1.0$ that correspond to curves from top to bottom.  The potential becomes flat at the transition point $q_c=\ln(2)/\ln(3)\simeq 0.63093$.}
  \label{V-m}
\end{figure}

To have a deeper understanding of the dynamics we can rewrite Eq.~(\ref{dmdt}) in the form of a time-dependent Ginzburg-Landau equation for $m$
\begin{equation}
  \frac{dm}{dt} = - \frac{\partial V(m)}{\partial m},
  \label{dmdt-1}
\end{equation}
where $V(m)$ is the associated potential.  This potential approach turns to be very useful to describe critical properties of systems with two symmetric absorbing states \cite{AlHammal-2005,Vazquez-2008-3}, as in the present model.  To obtain $V(m)$ we need to rewrite the right-hand side (rhs) of Eq.~(\ref{dmdt}) in terms of $m$.  Even though we do not know the exact relation between $n_1$ and $m$ we expect $n_1$ to be proportional to $m(1-m)$, given the fact that $n_1$ is zero at the extreme consensus states $m=0$ and $m=1$.  Then, using the Ansatz $n_1 \simeq c \, m(1-m)$, where $0 < c < 1$ is a constant, we obtain
\begin{equation}
  \frac{dm}{dt} = A_q \, m (1-m) \left( m-1/2 \right),
  \label{dmdt-2}
\end{equation}
with $A_q \equiv c \, (1-3P_{0,1})$.  Integrating the rhs of Eq.~(\ref{dmdt-2}) we arrive to the approximate Ginzburg-Landau potential 
\begin{equation}
  V(m) = - \frac{A_q}{8} \left( m-1/2 \right)^2 \left[1-2 \left( m-1/2 \right)^2  \right].
  \label{Vm}
\end{equation}
The shape of $V(m)$ from Eq.~(\ref{Vm}) is shown in Fig.~\ref{V-m} for four different values of $q$.  The single-well potential for $q<q_c$ becomes a double-well potential when $q>q_c$.  This picture is in agreement with the dynamics of the system that drives $m$ towards the minimum of $V(m)$, leading to a stationary state that corresponds to a coexistence of the three types of propensities for $q<q_c$ ($m=1/2$), and to consensus in $m=0$ or $m=1$ for $q>q_c$, both cases in a time that scales as $\ln N$.  The flat potential at $q_c$ (as in the voter model) denotes a purely diffusive dynamics of $m$ until it reaches an absorbing point $m=0$ or $m=1$ in a time that scales as $\tau \sim N$. \\

\subsection{Case $q=\infty$}

When $q=\infty$ the transition probabilities $P_{k,k'}$ from Eq.~(\ref{Ps}) are 
\begin{eqnarray*}
  P_{0,0}=P_{0,1}=P_{1,0} = 0, ~~ P_{1,1}=1/2, ~~ P_{1,2}=P_{2,1}=P_{2,2}=1,
\end{eqnarray*}
\begin{eqnarray*}
  P_{0,2} = 1-P_{2,0} = \begin{cases}
    1 ~~ & \mbox{if} ~~~ \omega<1/2, \\ 
1/2 ~~ & \mbox{if} ~~~ \omega=1/2, \\ 
0 ~~ & \mbox{if} ~~~ \omega>1/2. 
\end{cases}
\end{eqnarray*}
Notice that particles with state $k=1$ jump to either state $k=0$ or $k=2$ with the same probability $1/2$ when interacting with another $k=1$ particle, and are always attracted towards extreme states $k=0,2$ otherwise.  Particles with extreme states keep their states when they interact with particles with the same state or with $k=1$ particles.  Therefore, the behavior of the system is determined by the interactions between particles with opposite extreme states $k=0$ and $k=2$, which depends on the value of $\omega$.  For $\omega>1/2$ extreme particles never change state ($P_{0,k}=0$ and $P_{2,k}=0$ for $k=0,1,2$), and thus they attract $k=1$ particles until the system freezes in a complete bi-polarized state with no center particles $n_1^*=0$.  For $\omega<1/2$ extreme particles jump to state $k=1$ when they interact with opposite extreme particles, and thus we expect an active final bi-polarized state where the center state is non-empty ($n_1^*>0$).

The analysis developed in Appendix~\ref{S2_Stability_qinf} confirms the above expected behavior: 
\begin{itemize} 

\item[(i)] $\omega>1/2$: the fixed points are 
  \begin{equation*}
    \begin{split}
      & n_0^* = \frac{1}{2} \Big\{ 1+[n_0(0)-n_2(0)]\exp[n_1(0)] \Big\}, \\ 
      & n_1^* = 0, \\
      & n_2^* = \frac{1}{2} \Big\{ 1-[n_0(0)-n_2(0)]\exp[n_1(0)] \Big\}. 
    \end{split}
  \end{equation*}
  In particular, the mono-polarized consensus states $(1,0,0)$ and $(0,0,1)$ are obtained only when the system starts from that states, so that any initial condition different from mono-polarization leads to a bi-polarization.  
  
\item[(ii)] $\omega<1/2$:  the fixed points are the stable mono-polarized states $(n_0^*,n_2^*)=(1,0)$ and $(0,1)$ (sinks), and the unstable bi-polarized state  $n_0^*=n_2^*=(\sqrt{3}-1)/2$, $n_1^*=2-\sqrt{3}<n_0^*$ (saddle point).

\item[(iii)] $\omega=1/2$: same as for $\omega < 1/2$ but with an unstable bi-polarized state $n^*_0=n_2^*=\sqrt{2}-1$, $n_1^*=3-2\sqrt{2}<n_0^*$.  Notice that the bi-polarization in the $\omega=1/2$ case is stronger than that of the $\omega<1/2$ case given that $n_0^*(\omega=1/2) > n_0^*(\omega<1/2)$.

\end{itemize} 

In the next section we compare the analytical results of this section with that obtained by MC simulations of the model.

\section{Monte-Carlo simulations}
\label{MC}

\begin{figure}[t]
  \includegraphics[width=\columnwidth]{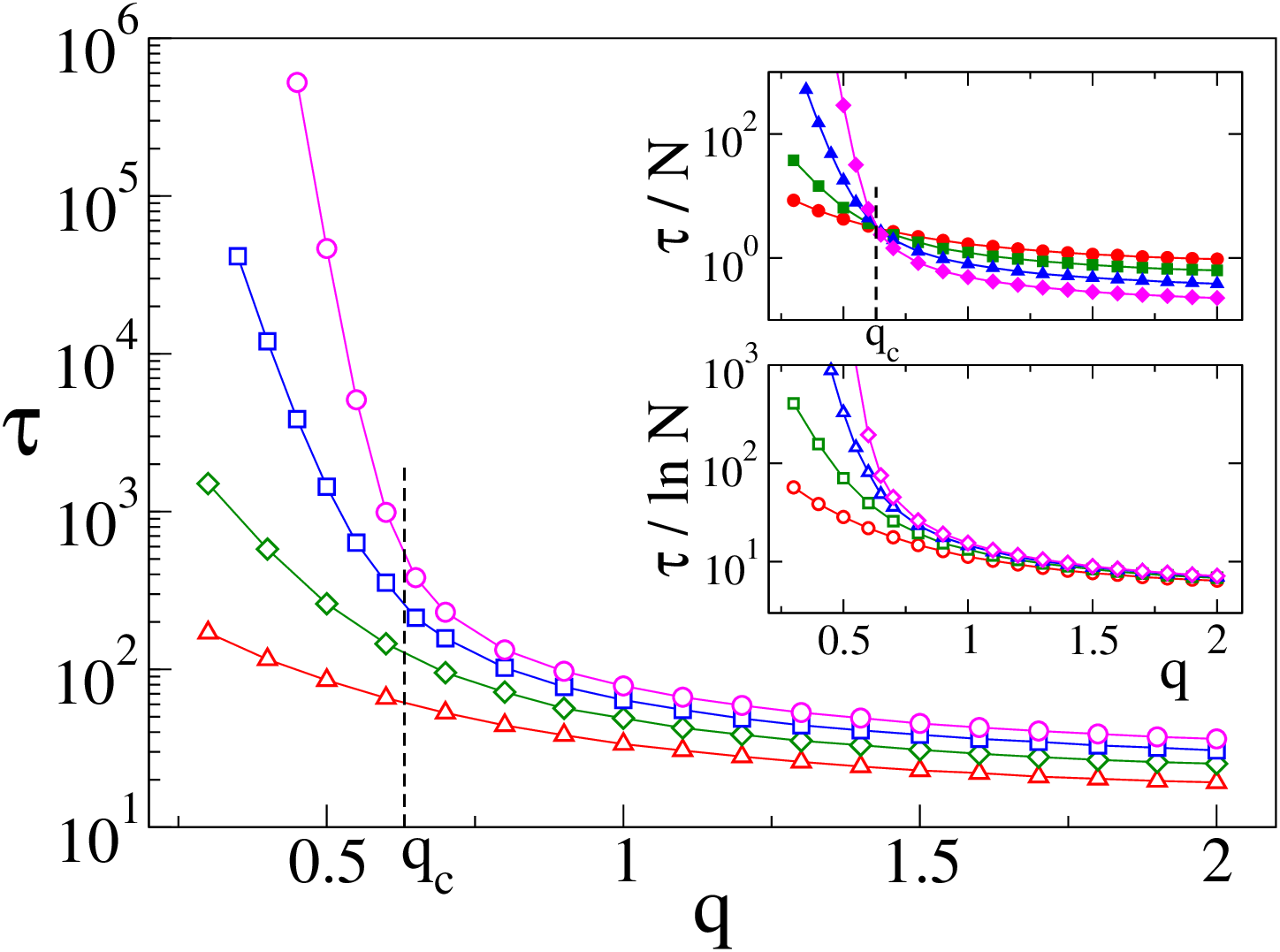}
  \caption{Mean consensus time $\tau$ vs $q$ for $S=2$, $\omega=0.5$, and system sizes $N=20$ (triangles), $N=40$ (diamonds), $N=80$ (squares), and $N=160$ (circles).  The vertical dashed line indicates the transition point $q_c (\omega=1/2) \simeq 0.631$.  Top inset: when $\tau$ is rescaled by $N$ all curves cross at $q_c$, where $\tau \sim N$.  Bottom inset: the collapse of the curves for large $q$ shows the scaling $\tau \sim \ln N$ for $q \gtrsim q_c$.}
  \label{tau-q}
\end{figure}

In this section we present MC simulation results for $S \ge 2$ and test the analytical predictions of section~\ref{S2} for $S=2$.  We run simulations of the model starting from a uniform distribution of propensities in $[0,1]$ and computed the mean consensus time $\tau$ as the average time to reach extreme consensus (one of the two mono-polarized states) over $10^4$ independent realizations of the dynamics.  We start by showing results for the simplest case $S=2$.  In Fig.~\ref{tau-q} we show $\tau$ vs $q$ from simulation results for $S=2$ and $\omega=0.5$.  Each of the four curves corresponds to a different system size $N$.  We observe that $\tau$ increases as $q$ decreases and becomes very large when $q$ falls below a value $q_c$ denoted by a vertical dashed line.  This abrupt increase is more clear for the largest system size $N=160$ (circles).  The reason for this is because $\tau$ increases very slowly (logarithmically) with $N$ when $q>q_c$ (see lower inset of Fig.~\ref{tau-q}) and very rapidly with $N$ (exponentially) when $q<q_c$ (not shown), while the growth seems to be linear with $N$ at $q_c$ (see upper inset of Fig.~\ref{tau-q}).  This behavior is reminiscent of a phase transition at $q_c$, where the scaling properties of the consensus time change.  Consensus is reached very fast for $q>q_c$ because once the initial symmetry in the propensity distribution is broken the system is quickly driven towards a mono-polarized state.  This is in agreement with the picture of a stable mono-polarization described in sections~\ref{stability} and \ref{potential} for $q>q_c$, where all particles move towards one of the extreme propensities ($p=0$ or $p=1$), driving the mean propensity $m$ towards the minimum of the potential at $m=0$ or $m=1$ (see Fig.~\ref{V-m}).  For $q<q_c$ the consensus is very slow given that the system falls into a stable bi-polarized state where $m$ fluctuates around the minimum of the potential at $m=1/2$ (Fig.~\ref{V-m}).  Eventually, one of the absorbing states at $m=0$ or $m=1$ is reached by a very large fluctuation in a time that is proportional to the exponential in the height of the potential or $N$.

\begin{figure}[t]
  \includegraphics[width=\columnwidth]{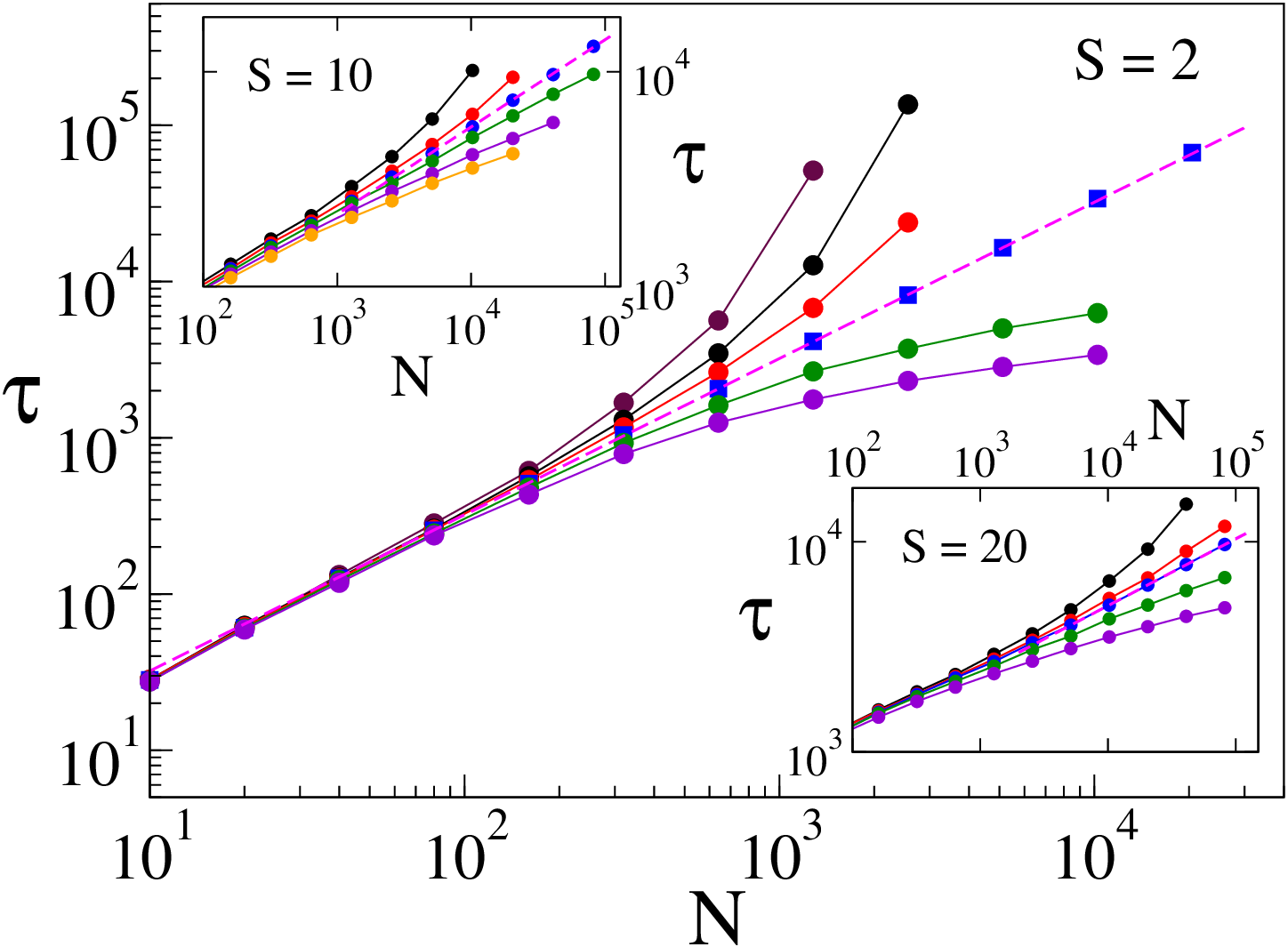}
  \caption{$\tau$ vs $N$ for $\omega=0.5$ and $S=2$ (main panel), $S=10$ (top inset) and $S=20$ (bottom inset).  Each curve corresponds to a different value of $q$.  Main panel: curves are for $q=0.620, 0.625, 0.628, 0.63093, 0.635$ and $0.640$ (from top to bottom).  Squares are for the theoretical transition value $q_c = \ln(2)/\ln(3) \simeq 0.63093$ [Eq.~(\ref{qc-0.5})] that shows the scaling $\tau \sim N$.  The dashed line has slope $1$.  Top inset: $q=0.312, 0.313, 0.31352, 0.314, 0.315$ and $0.316$ (top to bottom curves).  The dashed line has slope $0.406$.  Bottom inset: $q=0.2330, 0.2337, 0.2340, 0.2350, 0.2370$ and $0.2400$ (top to bottom curves).  The dashed line has slope $0.318$.}
  \label{tau-N-q}
\end{figure}

To study in more detail the scaling properties of the polarization transition at $q_c$ we show in Fig.~\ref{tau-N-q} the behavior of $\tau$ with $N$ for $\omega=0.5$ and values of $q$ around $q_c$.  We run simulations for four different values of $S$ ($S=2, 3, 10$ and $20$) but we only show three cases in Fig.~\ref{tau-N-q}.  In the main panel for $S=2$ we confirm that $\tau$ grows as a power law of $N$ ($\tau \sim N^{\alpha}$, with $\alpha=1.0$) at the theoretical transition point $q_c(S=2,\omega=0.5)=\ln(2)/\ln(3) \simeq 0.63093$ [Eq.~(\ref{qc-0.5})].  Interestingly, the non-trivial exponent $\alpha$ and the transition point $q_c$ depend on $S$, that is, $\alpha \simeq 0.5$ at $q_c \simeq 0.538$ for $S=3$ (not shown), $\alpha \simeq 0.406$ at $q_c \simeq 0.31352$ for $S=10$ (top inset), $\alpha \simeq 0.318$ at $q_c \simeq 0.234$ for $S=20$ (bottom inset).

To compare these numerical values of $q_c$ with that from the linear stability analysis theory, we estimated numerically the value of $q_c$ for $S=2, 3, 10$ and $20$ by following the stability approach done for $S=2$ in section~\ref{stability}.  That is, we linearized the system of Eqs.~(\ref{dndt}) around the mono-polarized state $(1,0,..,0)$ and found numerically the $S$ eigenvalues of the matrix ${\bf M}$ using \emph{Mathematica}.  It turns out that all eigenvalues are negative for $q>q_c$, but when $q<q_c$ one eigenvalue becomes positive and thus $(1,0,..,0)$ is unstable.  Therefore, we estimated $q_c$ by calculating the largest eigenvalue $\lambda_{\mbox{\scriptsize max}}(q)$ for decreasing values of $q$ with a resolution of $\Delta q=10^{-7}$, and determining $q_c$ as the first value of $q$ for which  $\lambda_{\mbox{\scriptsize max}}$ overcomes $0$, i e., $\lambda_{\mbox{\scriptsize max}}(q_c) \gtrsim 0$.  We obtained the transition values $q_c = 0.6309297, 0.538075, 0.3135217$ and $0.2336991$ for $S=2, 3, 10$ and $20$, respectively.  We can see that these values are very similar to that obtained from the scaling analysis of MC simulations (Fig.~\ref{tau-N-q})
 
\begin{figure}[t]
  \includegraphics[width=\columnwidth]{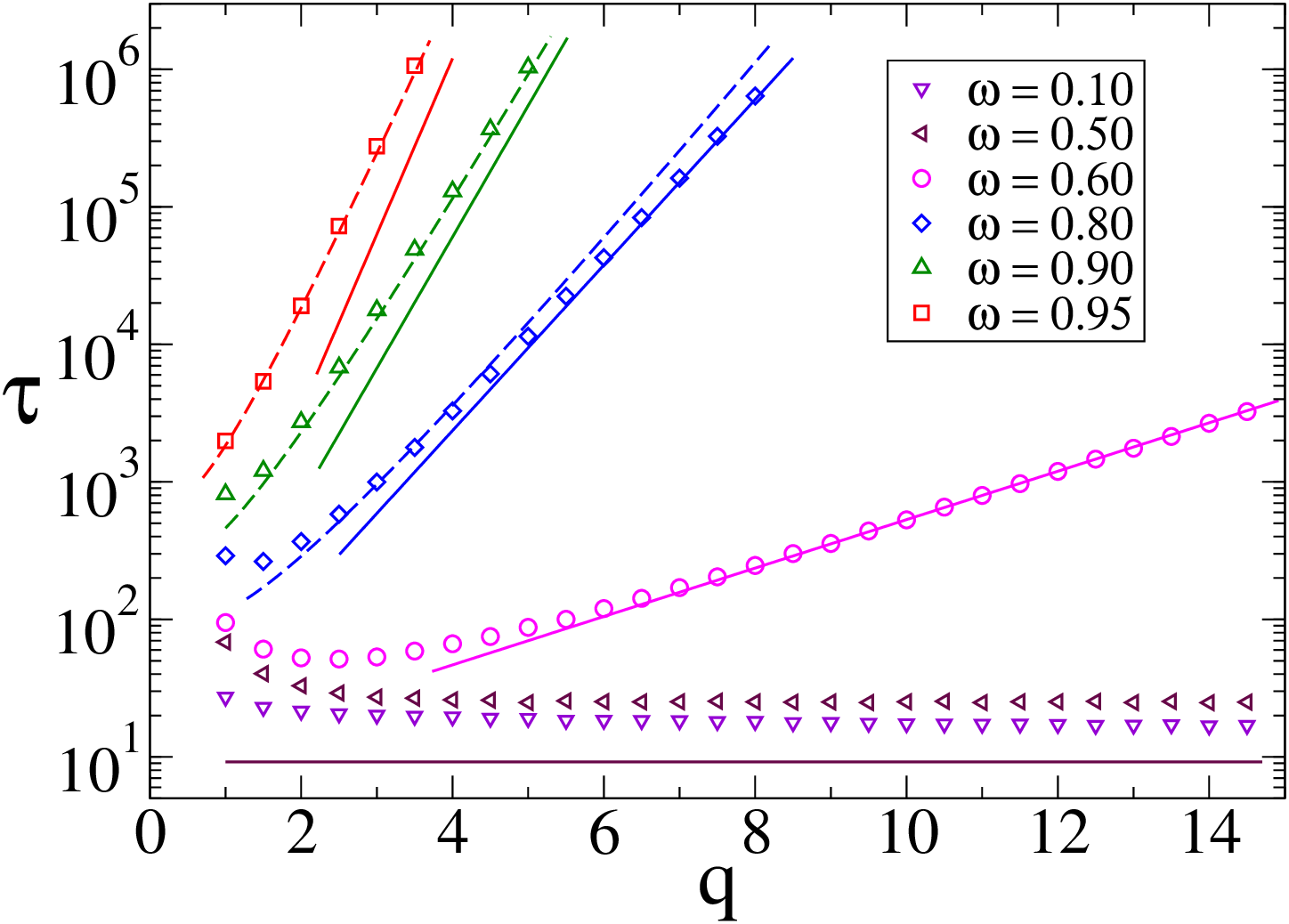}
  \caption{$\tau$ vs $q$ for $N=100$ and the values of $\omega$ indicated in the legend.  Solid lines correspond to the analytical approximation from Eq.~(\ref{tau-q-gg-1}) for $q \gg 1$, while dashed lines represent the analytical approximation from Eq.~(\ref{tau-q-g-1}) for $q>1$ and $\omega \to 1$.}    
  \label{tau-q-2}
\end{figure}

Figure~\ref{tau-q-2} shows the behavior of $\tau$ with $q>1$ for $N=100$ and several values of $\omega$.  We observe that $\tau$ diverges with $q$ for $\omega > 0.5$, and that it settles in a constant value for $\omega \le 0.5$, as suggested by the theoretical analysis of section~\ref{consensus}.  Solid lines represent the analytical prediction Eq.~(\ref{tau-q-gg-1}) for $q \gg 1$, which works well for large enough $q$ values that are visible for the
$\omega =0.6$ and $0.8$ curves, and gives a constant value proportional to $\ln(N)$ for $\omega \le 0.5$ that underestimates the numerical values for $\omega=0.5$ and $0.1$.  Dashed lines are the theoretical approximation from 
Eq.~(\ref{tau-q-g-1}) for $\omega \lesssim 1$, which fits the data points quite well for $\omega=0.9$ and $0.95$.  

\begin{figure}[t]
  \includegraphics[width=\columnwidth]{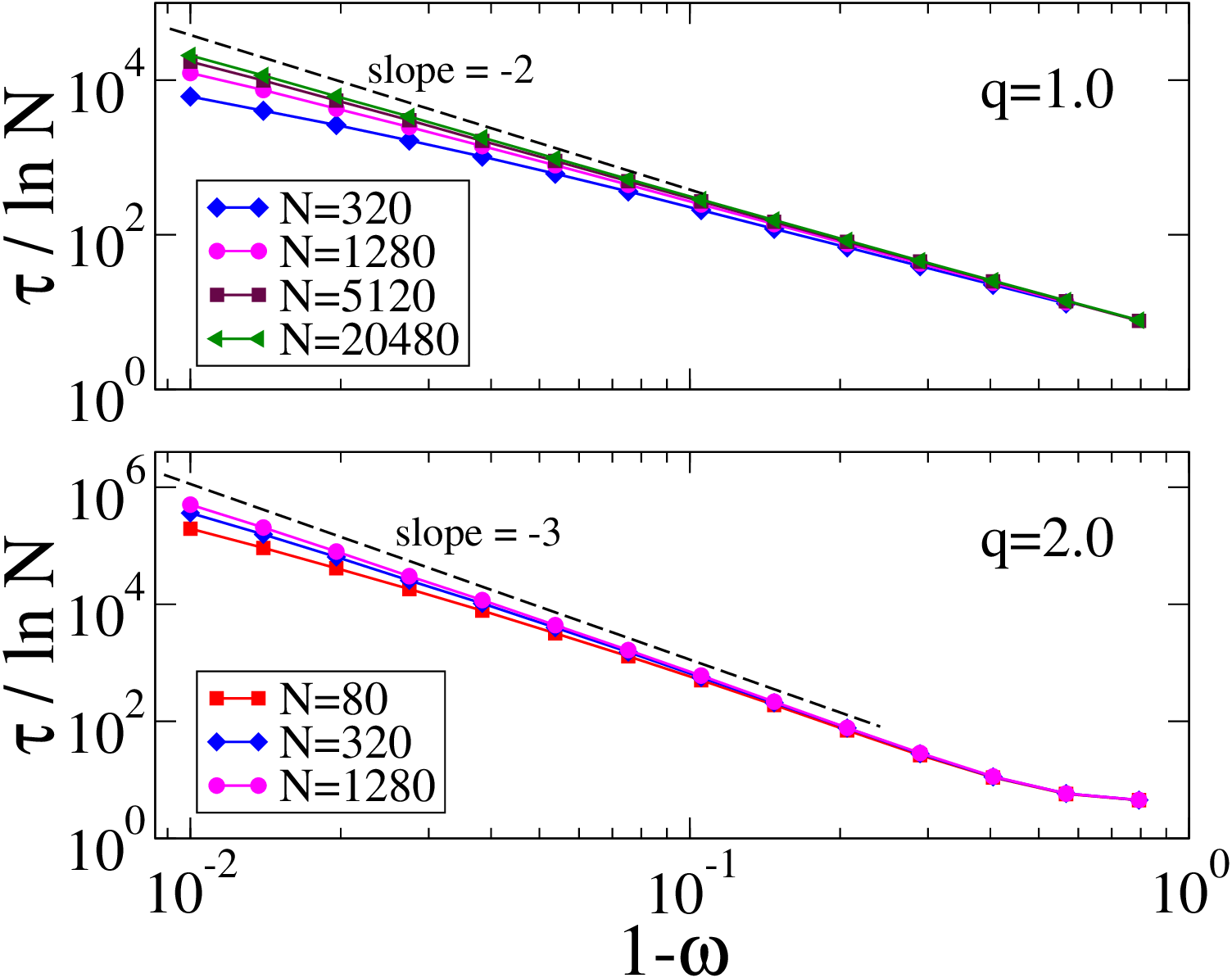}  
  \caption{$\tau$ vs $1-\omega$ for $S=2$, $q=1.0$ (top panel) and $q=2.0$ (bottom panel), and for the system sizes $N$ indicated in the legend.  The collapse of the data points for large $N$ shows that $\tau$ diverges as $(1-\omega)^{-(q+1)} \ln N$ when $\omega$  approaches $1.0$ [Eq.~(\ref{tau-q-g-1})].  Dashed lines have slopes $-(q+1)$.}
  \label{tau-w}
\end{figure}

To test the validity of Eq.~(\ref{tau-q-g-1}) which shows that $\tau$ diverges when $\omega$ approaches $1.0$, as we can also see in Fig.~\ref{tau-q-2} for a given $q$, we measured $\tau$ in MC simulations for $\omega$ very close to $1.0$.  Results are shown in Fig.~\ref{tau-w} for $q=1.0$ and $2.0$, top and bottom panels, respectively, where each curve corresponds to a different value of $N$ and the data was rescaled by $\ln N$ as suggested by Eq.~(\ref{tau-q-g-1}).  We observe that, for both $q$ values, the curves for large $N$ collapse into a single curve with a slope close to $-(q+1)$, in agreement with Eq.~(\ref{tau-q-g-1}).  For $q=2.0$, the slopes of $\tau$ vs $1-\omega$ for $\omega \lesssim 1$ seem to rapidly approach the theoretical value $-3$ as $N$ increases, but for $q=1.0$ the approach with $N$ to the slope $-2$ is much slower.  However, both cases seem to approach to the slope $-(q+1)$ in the thermodynamic limit. 

\begin{figure}[t]
  \includegraphics[width=\columnwidth]{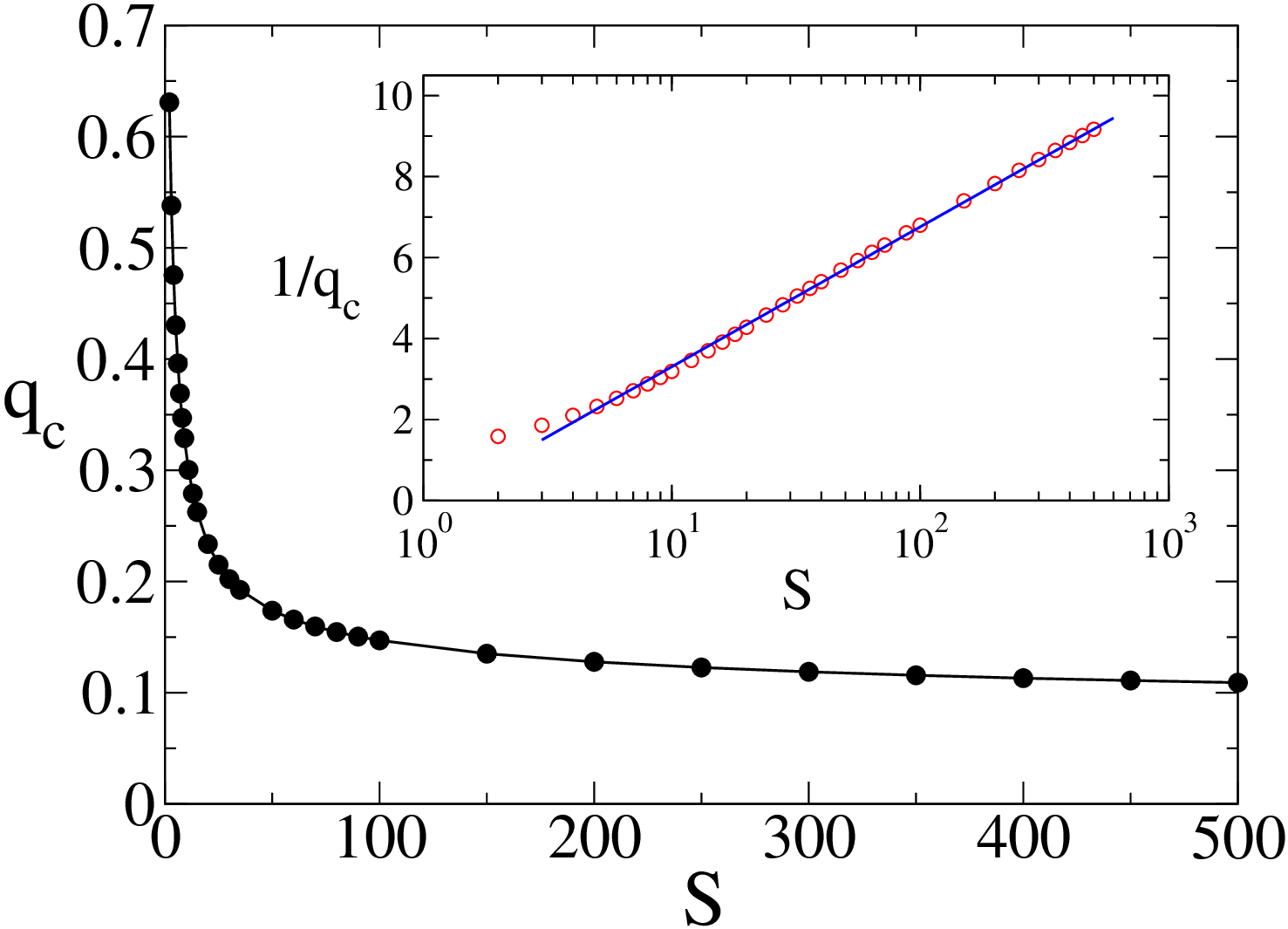}  
  \caption{Transition value $q_c$ vs $S$ for $\omega=0.5$ calculated numerically within the linear stability analysis of section~\ref{stability}.  At $q_c$ the largest eigenvalue of the linearized matrix ${\bf M}$ becomes $0$.  The solid line in the inset is the Ansatz $q_c = [1.5 \ln(S)-0.15]^{-1}$, showing that $q_c$ approaches very slowly to $0$ as $S$ increases.}
  \label{qc-S}
\end{figure}

The simulation results of Fig.~\ref{tau-N-q} show that $q_c$ seem to decrease as $S$ increases ($q_c \simeq 0.63093, 0.538, 0.31352$ and $0.234$ for $S=2, 3, 10$ and $20$, respectively), possibly suggesting that $q_c$ approaches zero as $S$ grows to infinity.  Given that MC simulations for each $S$ are very costly in terms of computational times, we estimated numerically $q_c$ for several values of $S$ using the linear stability approach of section~\ref{stability}, where the eigenvalues of the linearized matrix {\bf M} were calculated numerically.  Results are shown in Fig.~\ref{qc-S} for $\omega=0.5$, where we see that $q_c$ decreases very slowly with $S$.  The simple Ansatz
$q_c(\omega=0.5,S) \simeq [1.5 \ln (S)-0.15]^{-1}$ fits the data well for large $S$ (solid line in the inset of Fig.~\ref{qc-S}), showing that $q_c \to 0$ as $S \to \infty$.  However, due to the extremely slow decay of $q_c$ with $S$ it is hard to be sure whether $q_c$ reaches a constant value larger than zero or it eventually decays to zero.  To tackle this question we study the continuum limit $h \to 0$ ($S \to \infty$) in the next section, where we develop a theoretical approach that allows to deal with the system in the limit of continuum propensities in the real interval $[0,1]$.

\section{$S=\infty$ case: continuous approximation}
\label{continuous}

We showed in the last section that the mono-polarized consensus states are unstable for $q<q_c$ and stable for $q>q_c$, and that $q_c$ decreases monotonically with $S$, albeit very slowly.  In the $S \to \infty$ limit we thus expect the mono-polarized states to be unstable when $q \le 0$ and stable for any $q>0$, i e., $q_c(S = \infty)=0$.  To test this hypothesis we consider below the system of rate Eqs.~(\ref{dndt}) in the limiting case of a very small step $h \ll 1$ ($S \gg 1$).  This allows to derive a continuous in $p$ partial differential equation that describes the long-time behavior of the system.  We then linearize this equation around the mono-polarized state in order to study the stability of the system and find $q_c$.

\subsection{The continuous equation}

As explained in section~\ref{rate}, after a short initial transient all particles take discrete propensities $0,h,2h,..,1$, and thus the propensity distribution can be written as 
\begin{equation}
  f(p,t) = \sum_{k=0}^S n_k(t) \, \delta(p-kh), 
  \label{Measureh} 
\end{equation}
where $\delta(p-kh)$ is the Dirac delta function at $kh$.  Notice that the mono-polarized states correspond to $f(p)=\delta(p)$ and $f(p)=\delta(p-1)$.  
The mean value over the population of a generic function $\phi(p)$ of the propensity ("observable") is defined as
\begin{equation}
  \langle \phi \rangle_f(t) \equiv \int_0^1 \phi(p) f(p,t) dp =
  \sum_{k=0}^S n_k(t) \, \phi(kh). 
\end{equation}
For instance by taking $\phi(p)=p$ we obtain the mean propensity $m(t)=\langle p \rangle(t)$, and $\phi(p)=(p-\langle p \rangle)^2$ we obtain the variance of the propensity.  In Appendix~\ref{continuum-eq} we show that the time evolution of $\langle \phi \rangle_f$ is described by the following equation:
\begin{eqnarray}
  \label{dphidt}
  \frac{1}{h}\frac{d}{dt} \langle \phi \rangle_f &=& 
  \Big \langle V[f](p,t) \, \phi'(p) + \frac{h}{2}\phi''(p) \Big \rangle_f \nonumber \\
  &+& \Big\{ 1-B_0[f](t) \Big\} \, f(0,t) \left[ \phi'(0) - \frac{h}{2} \phi''(0) \right] \\
  &-& \Big\{ 1-B_1[f](t) \Big\} \, f(1,t) \left[ \phi'(1) + \frac{h}{2} \phi''(1) \right] + O(h^2), \nonumber 
\end{eqnarray}
where 
\begin{eqnarray}
  \label{Defv} 
  V[f](p,t) &\equiv& 2\int_0^1 \mathbf{P}^+(p,p')\,f(p',t) \, dp'-1, \\
  \label{DefB0} 
  B_0[f](t) &\equiv& \sum_j n_jP_{0,j} = \int_0^1 \mathbf{P}^+(0,p')\,f(p',t)dp' ~~ \mbox{and} \\
  \label{DefB1} 
  B_1[f](t) &\equiv& 1-\sum_j n_jP_{S,j} = 1-\int_0^1 \mathbf{P}^+(1,p')\,f(p',t)dp', \nonumber \\
\end{eqnarray}
with $\mathbf{P}^+(p,p')$ defined in Eq.~(\ref{Pij}).  The symbol "$[f]$" in front of $V$, $B_0$ and $B_1$ is used to denote that the integrals are with respect to the function $f \equiv f(p,t)$; a detailed notation that will be very useful when working with other functions.  For $q=1$ is $\mathbf P^+(p,p')=\omega p+(1-\omega)p'$, and thus we recover the corresponding coefficients 
\begin{eqnarray*}
  V[f](p,t) &\equiv& 2m(t)-1 + 2 \omega \left[ p-m(t) \right], \\
  B_0[f](t) &\equiv& (1-\omega) m(t) ~~ \mbox{and} \\
  B_1[f](t) &\equiv& (1-\omega) \left[ 1-m(t) \right]
\end{eqnarray*}
of the linear model studied in \cite{Vazquez-2020}.

As we can see in the derivation of Appendix~\ref{continuum-eq}, there are no $O(h^2)$ terms when $\phi$ is linear in $p$.  Moreover the first term in the rhs of Eq.~(\ref{dphidt}) comes from the rate equations for $n_k(t)$ ($0<k<S$) that describe the evolution of the propensity distribution in $(0,1)$, while the second and third terms come from the dynamics near the boundary points at $p=0$ and $p=1$, respectively, and describe the balance between the particles entering and leaving the boundary.  
Here $B_0[f](t)$  and $B_1[f](t)$ are boundary coefficients, while $V[f](p,t)=\langle \mathbf P^+ \rangle - \langle \mathbf P^- \rangle$ is the average jumping probability of a particle with propensity $p$ at time $t$, and gives the net drift towards the ends of the interval $[0,1]$, so that the particle would tend to move to the right when $V[f](p,t)>0$ and to the left when $V[f](p,t)<0$.

Taking $\phi(p)=1$ in Eq.~(\ref{dphidt}) leads to the conservation of the total mass $\int_0^1 f(p,t) dp=1$, as expected.  Besides, for $\phi(p)=p$ we obtain the following equation for the evolution of the mean propensity:
\begin{eqnarray*}
  \frac{1}{h}\frac{d}{dt}m(t) 
  &=& \int_0^1 V[f](p,t)\,f(p,t) \, dp + f(0,t) \Big\{ 1-B_0[f](t) \Big\} \\ 
  && - f(1,t) \Big\{ 1-B_1[f](t) \Big\}.
\end{eqnarray*}
We can check from this expression that if $m(0)=0$ [$m(0)=1$] then $m(t)=0$ [$m(t)=1$] for all $t \ge 0$ since the rhs is zero when $f=\delta(p)$ [$f=\delta(p-1)$]. Thus the mono-polarized states are stationary states as expected.  

To better explore the dynamics we derive below an approximate equation for the time evolution of the propensity distribution $f(p,t)$.  For that, we rewrite Eq.~(\ref{dphidt}) neglecting order $h$ terms as
\begin{equation}
  \label{Equ100}
  \frac{1}{h} \frac{d}{dt} \langle\phi\rangle_f = \int_0^1 \big\{ V[f](p,t)+u[f](p,t) \big\} f(p,t)\phi'(p)\,dp,\
\end{equation}
where we have introduced the field 
\begin{eqnarray*}
  u[f](p,t) = \big\{ 1-B_0[f](t) \big\} \delta(p) - \big\{ 1-B_1[f](t) \big\} \delta (p-1).
\end{eqnarray*}
Integrating by parts the rhs. of Eq.~(\ref{Equ100}) and regrouping terms leads to 
\begin{eqnarray*}
  && \int_0^1 \phi(p) \bigg\{  \frac{1}{h} \frac{\partial}{\partial t} f(p,t) 
  + \frac{\partial}{\partial p} \Big\{ \left[ V[f](p,t)+u[f](p,t) \right] f(p,t) \Big\} \bigg\} dp \\ 
  && = \big\{ V[f](1,t)+u[f](1,t) \big\} f(1,t) \phi(1) \\
  && - \big\{ V[f](0,t)+u[f](0,t) \big\} f(0,t) \phi(0). 
\end{eqnarray*}
Since this relation holds for any function $\phi(p)$, we see that $f$ satisfies formally the transport equation 
\begin{eqnarray}
  && \frac{1}{h}\frac{\partial}{\partial t} f(p,t) 
  + \frac{\partial}{\partial p} \Big\{ \left[ V[f](p,t)+u[f](p) \right] f(p,t) \Big\}  \nonumber \\
  && = \left\{ V[f](1,t)+u[f](1,t) \right\} f(1,t)\delta(p-1) \nonumber \\
  && - \left\{ V[f](0,t)+u[f](0,t) \right\} f(0,t) \delta(p).
  \label{dfdt} 
\end{eqnarray}
This equation expresses the conservation of the total number of particles under the transport induced by the effective drift $V+u$ and with source terms
$h \left[ V[f](1,t)+u[f](1,t) \right] f(1,t) \delta(p-1)$ and $- h \left[ V[f](0,t)+u[f](0,t) \right] f(0,t)\delta(p)$ at the boundary points $p=1$ and $p=0$, respectively.  An intuitive interpretation of this equation is that the mass density $f(p,t)$ is transported by the field $V[f](p,t)$ in $[0,1]$ and suffers an additional impulse at the borders $p=0$ and $p=1$ given by the field $u[f](p,t)$, which is associated to the rate Eqs.~(\ref{dndt0}) and (\ref{dndtS}) for $n_0$ and $n_S$, respectively.

\subsection{Stability of the mono-polarized states}
\label{stability-delta}

Here we study the stability of the mono-polarized state $\delta(p)$ with the aim of finding the transition point $q_c$.  We consider an initial small perturbation of $\delta(p)$ of the form $\varepsilon(0) \gamma(p,0)$, where $0 \le \varepsilon(0) \ll 1$ is the size of the perturbation and $\gamma(p,0)$ is a probability measure supported in $[0,\Delta]$ for some small $\Delta>0$ such that $\gamma(p=0,0)=0$ and $\int_0^1 \gamma(p,0) dp=1$.  Thus, the initial propensity distribution can be written as $f(p,t=0)=[1-\varepsilon(0)] \delta(p) + \varepsilon(0) \gamma(p,0)$.  In other words, we are removing a fraction $\varepsilon(0)$ of particles from $p=0$ and redistribute them in the interval $[0,\Delta]$ following a generic function $\gamma(p,0)$.  We can then write the corresponding solution $f(p,t)$ of Eq.~(\ref{dphidt}) in the form 
\begin{equation}
	f(p,t)=[1-\varepsilon(t)] \delta(p) + \varepsilon(t) \gamma(p,t), 
	\label{fpt}
\end{equation}	
where $0 \le \varepsilon(t) \le 1$ and $\gamma(p,t)$ is a probability measure such that 
$\gamma(p=0,t)=0$ and $\int_0^1 \gamma(p,t) dp=1$ for all $t \ge 0$. 

We want to find the conditions on $q$ such that $f(p,t) \to \delta(p)$ as $t \to \infty$, and thus $\delta(p)$ is locally asymptotically stable.  Inserting the expression from Eq.~(\ref{fpt}) for $f(p,t)$ into Eqs.~(\ref{Defv})-(\ref{DefB1}) we obtain
\begin{eqnarray*}
        V[f](p,t) & = & V[\delta](p) \\		
	&+& 2\varepsilon(t) \left[ \int_0^1 \mathbf{P}^+(p,p') \, \gamma(p',t) \, dp' -\mathbf{P}^+(p,0) \right],   \\ 
	B_0[f](t) & = &   \varepsilon(t) B_0[\gamma](t), \\ 
	B_1[f](t) & = & B_1[\delta](t) \\ &+& \varepsilon(t) \left[ \mathbf{P}^+(1,0) 
	- \int_0^1 \mathbf{P}^+(1,p') \, \gamma(p',t) \, dp'   \right],
\end{eqnarray*} 
where $\delta \equiv \delta(p)$ and we have used $\mathbf{P}^+(0,0)=0$.  Plugging the above expressions for $V$, $B_0$ and $B_1$ into Eq.~(\ref{dphidt}) and neglecting terms of order $\varepsilon^2 \ll 1$ and $h^2 \ll 1$ we obtain, after doing some algebra, the linearized equation 
\begin{eqnarray}
	\label{Equ11} 
	\frac{1}{h}\frac{d}{dt} \langle \phi \rangle_{f} 
	& = &  \varepsilon(t)  \Big\langle V[\delta](p) \, \phi'(p) \Big \rangle_{\gamma} \\
	& + & \varepsilon(t) \phi'(0) B_0[\gamma](t)  + \varepsilon(t) \big\{ 1-B_1[\gamma](t) \big\} \gamma(1,t) \phi'(1), \nonumber
\end{eqnarray}
which holds for any continuous function $\phi(p)$.  The last two terms in the rhs of Eq.~(\ref{Equ11}) account for particles with propensity $0$ or $1$ that bounce back into the interval $(0,1)$.  Taking in particular $\phi$ with compact support in the interval $(0,1)$ we obtain  the following weak form of the transport equation:
\begin{equation} 
	\frac{1}{h}\frac{\partial}{\partial t} \big\{ \varepsilon(t) \, \gamma(p,t) \big\}
	+ \frac{\partial}{\partial p} \big\{ \varepsilon(t) \, V[\delta](p)  \, \gamma(p,t)  \big\} = 0. 
	\label{dgammadt}
\end{equation}
Equation~(\ref{dgammadt}) describes the evolution of the distribution $\gamma(p,t)$ of particles with propensity in $(0,1)$, which are transported by the vector-field $V[\delta](p)$ generated by $0$--propensity particles.  That is, to first order in $\epsilon(t) \ll 1$, a particle with propensity $p>0$ interacts most likely with particles with propensity $p=0$, and thus it "feels" a field $V[\delta](p)$ that tends to move the particle to the right if $V[\delta](p)>0$ and to the left is $V[\delta](p)<0$.  This result is very important because if it happens that the drift $V[\delta](p)$ is negative for all $p$ in the support $[0,\Delta]$ of $\gamma(t)$, then all particles would be attracted towards $p=0$, and thus $\delta(p)$ would be stable.  As we show below, $V[\delta](p)$ becomes negative when $q$ becomes positive, under the condition $p<(2\omega)^{-1}$ (see Appendix~\ref{stability-delta-proof} where we relax this condition and give a more general proof of the results presented below).  Indeed, for a given $\omega$ we assume that $\Delta < (2 \omega)^{-1}$ and thus all particles have propensity $p<(2 \omega)^{-1}$.  Then, since
\begin{eqnarray*} 
	V[\delta](p)=2 \mathbf{P}^+(p,0)-1 = \frac{2(\omega p)^q}{(\omega p)^q+(1-\omega p)^q} - 1
\end{eqnarray*}	 
and $\omega p < 1/2$, we can check that 
\begin{eqnarray*}
	&& V[\delta](p)<0 ~~ \mbox{for $q>0$,  ~ and} \\  
	&& V[\delta](p)>0 ~~ \mbox{for $q<0$}.  
\end{eqnarray*}
Therefore, the linear stability analysis shows that the mono-polarized state $\delta(p)$ is unstable for $q<0$ and locally asymptotically stable for $q>0$, for the linearized Eq.~(\ref{Equ11}).  Although we understand that there is no linear stability theory available for partial differential equations like Eq.~(\ref{dfdt}) akin to that of the classical Dynamical Systems, these findings, along with the results obtained for finite $S$, strongly suggest that $\delta(p)$ is unstable for $q<0$ and stable for $q>0$. 

In summary, in this section we showed that in the continuum limit $S \to \infty$ the polarization transition happens at a threshold value $q_c(S=\infty)=0$.

\section{Summary and discussion}
\label{summary}

In this paper we investigated a generic nonlinear version of the model introduced in \cite{Vazquez-2020} that studies the dynamics of voting intention in a interacting population of agents.  Each agent has a propensity in the interval $[0,1]$ to vote for one of two candidates that can either increase or decrease after interacting with another agent.  We considered the general case scenario in which the probability to increase or decrease the propensity is a nonlinear function of the weighted average propensity of the two interacting agents, whose shape is controlled by a parameter $q \ge 0$.  We studied by MC simulations and a rate equation approach the stability properties of the stationary states of the system (mono-polarization and bi-polarization), as well as the time to reach the mono-polarized state.  In particular, we explored the conditions for which the bi-polarized state is stable.  In the linear ($q=1$) version of the model studied in \cite{Vazquez-2020}, the evolution of the system exhibits two stages starting from a uniform distribution of propensities: an initial quick approach to a bi-polarized state where the propensity distribution is peaked at the extreme values $0$ and $1$, followed by a relaxation to a mono-polarized state, that is, a consensus in an extreme propensity.  Bi-polarization in this linear case is unstable given that, once the symmetry of the system is broken by finite-size fluctuations, the population is driven to an extreme consensus.  This transient state of bi-polarization has a short lifetime that increases very slowly (logarithmically) with the system size $N$.

The nonlinear update rule studied in this article is able to generate a bi-polarized state that is stable when $q$ is smaller than a threshold value $q_c$, while the mono-polarized state is stable for $q>q_c$.  An intuitive explanation of this behavior can be given in terms of a competition between two different mechanisms that can be better understood if we see the dynamics of agents' propensities as particles that jump to their neighboring sites in a one-dimensional chain in the interval $[0,1]$.  One mechanism is a drift that induces a tendency in the particles to move in the direction of the mean propensity, which generates a positive feedback that breaks the symmetry of the system and leads all particles to an extreme consensus.  The other mechanism is a ``sticky'' border effect by which a particle that reaches an extreme propensity value $0$ or $1$ tends to remain there, attracting other particles to that border.  This effect generates two opposite poles of particles that tend to remain stable.  For low $q$ values ($q<q_c$) particles jump right and left with probabilities similar to $1/2$, performing a nearly symmetric diffusion in the interval $(0,1)$.  Therefore, the drift towards the mean propensity is weak and thus the sticky border effect dominates, inducing a stable bi-polarization.  For $q>q_c$ the drift becomes intense enough to overcome the sticky effect, breaking the stability of the bi-polarized state and leading the system to a stable mono-polarization.  The absorbing extremist consensus is eventually achieved in a finite system by demographic fluctuations, in a mean time that grows exponentially fast with $N$ for $q<q_c$ and logarithmically with $N$ for $q>q_c$.  At the transition point $q_c$, the mean consensus time shows the power law scaling $\tau \sim N^{\alpha}$, where $\alpha(h)$ is an $h$-dependent non-universal exponent.  Interestingly, $q_c$ decreases and seem to vanish as $h$ approaches $0$.  An insight into this result was obtained by studying a transport equation derived in the continuum $h \to 0$ limit, which shows that, indeed, the transition between mono-polarization and bi-polarization is at $q_c(S \to \infty)=0$.

It would be worthwhile to study versions of this model where pairwise interactions are not simply taken as all-to-all, but rather take place on lattices or complex networks.  It might also be interesting to study the stability of bi-polarization under the presence of an external noise that allows agents to choose propensities at random.

\begin{acknowledgments}
This work was partially supported by Universidad de Buenos Aires under grants 20020170100445BA, 20020170200256BA, and by ANPCyT PICT2012 0153 and PICT2014-1771.  
\end{acknowledgments}

\section*{AIP Publishing Data Sharing Policy}

The data that supports the findings of this study are available within the article.

\onecolumngrid

\appendix

\section{Fixed points for the $S=2$ case and $q=\infty$}
\label{S2_Stability_qinf}

Assume that $\omega>1/2$.  The rate Eqs.~(\ref{dndt-2}) become 
\begin{eqnarray*}
	\frac{dn_0}{dt} &=& \frac12(1-n_0-n_2)(1+n_0-n_2) \\ 
	\frac{dn_2}{dt} &=& \frac12(1-n_0-n_2)(1-n_0+n_2).  
\end{eqnarray*}
We shall see below that the stationary solution corresponds to a distribution with no particles with state $k=1$ ($n_1^*=0$).  Notice that $n_0$ and $n_2$ are nondecreasing functions thus $n_0(t)\ge n_0(0)$ and $n_2(t)\ge n_2(0)$. 
Moreover $n_0+n_2$ and $n_0-n_2$ satisfy
\begin{eqnarray*}
	&& \frac{d}{dt}(n_0+n_2) = 1-(n_0+n_2), \\
	&& \frac{d}{dt}(n_0-n_2) = [1-(n_0+n_2)](n_0-n_2). 
\end{eqnarray*}
The solution of these equations is
\begin{eqnarray*}
  n_0(t)&=& \frac{1}{2} \Big\{ 1-n_1(0)e^{-t}+[n_0(0)-n_2(0)] \exp[ n_1(0)(1-e^{-t})] \Big\}, \\
  n_2(t)&=& \frac{1}{2} \Big\{ 1-n_1(0)e^{-t}-[n_0(0)-n_2(0)] \exp[n_1(0)(1-e^{-t})] \Big\}. 
\end{eqnarray*}
Then, in the long time limit we have
\begin{equation}
  \label{S2_qinf}
\begin{split}
  n_0(t) \to n_0^* = \frac{1}{2} \Big\{ 1+[n_0(0)-n_2(0)] \exp[n_1(0)] \Big\}, \\ 
  n_2(t) \to n_2^* = \frac{1}{2} \Big\{ 1-[n_0(0)-n_2(0)] \exp[n_1(0)] \Big\}. 
\end{split}
\end{equation}
The extreme consensus state $(n_0^*,n_2^*)=(1,0)$ happens when 
$[n_0(0)-n_2(0)] \exp[n_1(0)]=1$, thus we need $n_2(0)=0$ 
[since $n_2(t)\ge n_2(0)$] and so then $n_0(0) \exp[1-n_0(0)]=1$, that is, $n_0(0)=1$.  The same argument shows that the extreme consensus state $(n_0^*,n_2^*)=(0,1)$ is obtained only when starting from that consensus state $(0,1)$.  Therefore, we conclude that when $q=\infty$ and $\omega>1/2$ any initial condition which is not a consensus state will asymptotically lead to a bi-polarized state $(n_0^*,n_2^*)$ with $n_0^*,n_2^*>0$, $n_0^*+n_2^*=1$ ($n_1^*=0$) as given by Eq.~(\ref{S2_qinf}), where there are no particles with state $k=0$. \\

For $\omega<1/2$ the rate Eqs.~(\ref{dndt-2}) are 
\begin{equation}\label{S2_qinf_rate}
\begin{split}
& \frac{d}{dt}n_0 = n_0n_1+\frac12 n_1^2 - n_0n_2, \\ 
& \frac{d}{dt}n_2 = n_2n_1+\frac12 n_1^2 - n_0n_2. 
\end{split}
\end{equation}
To obtain the stationary solutions we subtract add and subtract both equations to get:
\begin{eqnarray*}
\begin{split}
& 0=n_1^*(n_0^*+n_2^*)+n_1^{*^2}-2n_0^*n_2^*, \\ 
& 0=(n_0^*-n_2^*)n_1^*. 
\end{split}
\end{eqnarray*}
The solutions of these equations are the mono-polarized states $(1,0,0)$ and $(0,0,1)$ and the bi-polarized state $n^*_0=n_2^*=(\sqrt{3}-1)/2$, $n_1^*=2-\sqrt{3}<n_0^*$.  Linearizing Eqs.~(\ref{S2_qinf_rate}) around these fixed points we find that the mono-polarized states are sinks and the bi-polarized state is a saddle.  Let $\mathcal{T}$ be the triangle $\{(n_0,n_2)\in \mathbb{R}^2,\, n_0,n_2\ge 0,\,n_0+n_2\le 1\}$. 
Denote $F(n_0,n_2)$ the vector-field  defined by the r.h.s of Eqs.~(\ref{S2_qinf_rate}).  When evaluated on the boundary of $\mathcal{T}$, $F$ points strictly inside $\mathcal{T}$.  Its divergence is $\text{div}\,F(n_0,n_2)=-2(n_0+n_2)$ which is negative inside $\mathcal{T}$
so that there is no periodic orbit inside $\mathcal{T}$ according to Bendixson Criteria (\cite{Perko}[section 3.9]).  Eventually since the consensus states are sinks and the polarized state $(n_0^*,n_2^*)$ is a saddle, the only possible separatrix cycle (\cite{Perko}[section 3.7]) is an homoclinic orbit connecting the stable and unstable manifold of the saddle.  Since $n_0^*=n_2^*$, the function $n_0-n_2$ when evaluated along this orbit reaches at some time $t^*$ an extremum which does not belong to the line $n_0=n_2$. 
Since $(n_0-n_2)'=n_1(n_0-n_2)$, we have $n_1(t^*)=0$ and thus the orbit touches the boundary $n_0+n_2=1$.  This is impossible since  the vector-field $F$ points strictly inside $\mathcal{T}$ along the boundary of $\mathcal{T}$. 
Thus there is no periodic orbit and no separatrix cycle inside $\mathcal{T}$. 
We then conclude with Poincare-Bendixson Theorem (\cite{Perko}[section 3.7]) that any trajectory of Eq.~(\ref{S2_qinf_rate}) must converge to an equilibrium point. Thus any trajectory converge to one of the consensus states except those lying in the stable manifold of the saddle (such initial conditions forms a set with zero measure). \\

When $\omega=1/2$ the rate Eqs.~(\ref{dndt-2}) are 
\begin{equation*}
\begin{split}
& \frac{d}{dt}n_0 = n_0n_1+\frac12 n_1^2 - \frac12 n_0n_2, \\ 
& \frac{d}{dt}n_2 = n_2n_1+\frac12 n_1^2 - \frac12  n_0n_2. 
\end{split}
\end{equation*}
We can repeat the same analysis as for the $\omega<1/2$ case and obtain that the fixed points are the mono-polarized states and the bi-polarized state $n^*_0=n_2^*=\sqrt{2}-1$, $n_1^*=3-2\sqrt{2}<n_0^*$.

\section{Consensus times for $S=2$: two limiting cases}
\label{stability-2}

As explained in section~\ref{consensus}, the mean consensus time is related to the largest eigenvalue $\lambda_{\mbox{\scriptsize max}}$ of matrix ${\bf A}$ by the expression $\tau \sim - \ln N/\lambda_{\mbox{\scriptsize max}}$, where $\lambda_{\mbox{\scriptsize max}}$ is the largest of $\lambda_\pm <0$.  The eigenvalues $\lambda_\pm$ are the roots of the characteristic polynomial of ${\bf A}$, $x^2-Tr({\bf A}) x + det({\bf A})$, where $Tr({\bf A})=P_{0,1}-P_{0,2}-1$ and $det({\bf A})=P_{0,2}(1-2P_{1,0}-P_{0,1})$ are the trace and the determinant of ${\bf A}$, respectively.  Below we analyze two special limits where we approximate $P_{0,2}$, $P_{1,0}$ and $P_{0,1}$ by doing a Taylor series expansion, and find analytical expressions for $\lambda_\pm = \frac12 \left[ Tr({\bf A}) \pm \sqrt{\Delta} \right]$, with  $\Delta=Tr^2({\bf A})-4 \, det({\bf A})$.

\subsection{Case $q\gg 1$, $\omega\in (0,1)$}
\label{wgg1}

We take $\omega\in (0,1)$ and define 
$$a \equiv \frac{1-\omega}{1+\omega},\qquad b \equiv \frac{\omega}{1-\omega},\qquad c \equiv \frac{1-\omega}{\omega},
\qquad d \equiv \frac{\omega}{2-\omega}.$$ 
Notice that $a,d\in (0,1)$ and that $b=1/c$ is less than $1$ only if $\omega<1/2$.  Then, for $q\gg 1$ we can approximate the expressions for $P_{k,k'}$ from Eq.~(\ref{Ps}) as 
$$P_{0,1}=a^q+O(a^{2q}),\qquad P_{1,0}=d^q+O(d^{2q}), \qquad 
P_{0,2} = \begin{cases} 1-b^q+O(b^{2q}) & ~~ \text{if $\omega<1/2$,} \\ 
c^q + O(c^{2q}) & ~~ \text{if $\omega>1/2$.}
\end{cases}
$$ 
Since $a<c$, so that $a^q\ll c^q$ when $q\gg 1$, it follows that 
\begin{eqnarray}
Tr({\bf A})= 
\begin{cases}
-2+a^q+b^q + O(a^{2q}+b^{2q}) & ~~ \text{if $\omega<1/2$}, \\ 
-1-c^q + o(c^q) & ~~ \text{if $\omega>1/2$,}
\end{cases}
\end{eqnarray}
and 
\begin{eqnarray}
det({\bf A})=
\begin{cases}
1-2d^q-a^q-b^q + o(a^q+b^q+d^q) & ~~ \text{if  $\omega<1/2$}, \\ 
c^q + o(c^q) & ~~ \text{if $\omega>1/2$.}
\end{cases}
\end{eqnarray}
Then
\begin{eqnarray}
  \sqrt{\Delta}
  =\begin{cases}
  2\sqrt{2}d^{q/2} + o(a^q+b^q+d^q) & ~~ \text{if $\omega<1/2$}, \\ 
  1-c^q+o(c^q) & ~~ \text{if $\omega>1/2$. } 
  \end{cases}
\end{eqnarray}
Therefore, we find that the eigenvalues of ${\bf A}$ are  
\begin{eqnarray}
  \lambda_\pm \simeq -1 ~~ && \mbox{if $\omega<1/2$} ~~ \mbox{and} \\
  \lambda_+ \simeq -c^q > \lambda_- \simeq -1 ~~ && \mbox{if $\omega>1/2$.}
\end{eqnarray}
Finally, the consensus time should scale as $\tau \sim \ln N$ if $\omega<1/2$ and as $\tau \sim \left( \frac{\omega}{1-\omega} \right)^q \ln N$ if $\omega>1/2$, as quoted in Eq.~(\ref{tau-q-gg-1}) of the main text.

\subsection{Case $1-\omega \ll 1$, $q \ge 1$}
\label{1-w-1}

Let us consider that $q=1$ and define $\epsilon \equiv 1-\omega \ll 1$, thus we have $P_{0,1}= \frac{\epsilon}{2}$, $P_{1,0}=\frac{1-\epsilon}{2}$ and $P_{0,2}=\epsilon$.  Then $Tr(A)=-1-\frac{\epsilon}{2}$, $\det({\bf A})=\frac{\epsilon^2}{2}$, and $\sqrt{\Delta}=1+\frac{\epsilon}{2}-\epsilon^2+ O(\epsilon^2)$.  We then find that $\lambda_- \simeq -1$ and $\lambda_+ \simeq -\frac{\epsilon^2}{2}$, and thus the consensus time should scale as $\tau \simeq \frac{2\ln\,N}{(1-\omega)^2}$. \\

Let us now take $q>1$, thus we can approximate $P_{k,k'}$ as 
\begin{equation}
  \label{AproxPij_omega1}
\begin{split}
P_{0,1} & = \frac{\epsilon^q}{\epsilon^q+(2-\epsilon)^q} = \frac{1}{2^q} \epsilon^q \left[ 1+\frac{q}{2} \epsilon+o(\epsilon) \right], \\ 
P_{1,0} & = \frac{(1-\epsilon)^q}{(1-\epsilon)^q+(1+\epsilon)^q} = \frac12 -\frac{q}{2}\epsilon+O(\epsilon^2), \\ 
P_{0,2} &  = \frac{\epsilon^q}{\epsilon^q+(1-\epsilon)^q} 
= \epsilon^q \left[ 1+q\epsilon+o(\epsilon) \right]. 
\end{split}
\end{equation} 
Then 
\begin{eqnarray*}
  Tr({\bf A}) & = &  -1-(1-1/2^q)\epsilon^q-q(1-1/2^{q+1})\epsilon^{q+1}+o(\epsilon^{q+1}),  \\
  det({\bf A}) &= &  q\epsilon^{q+1}+o(\epsilon^{q+1}) ~~ \mbox{and} \\ 
  \sqrt{\Delta} & = & 1+(1-1/2^q)\epsilon^q-q(1+1/2^{q+1})\epsilon^{q+1}+o(\epsilon^{q+1}).   
\end{eqnarray*}
The eigenvalues of ${\bf A}$ are then 
$\lambda_+ \simeq -q\epsilon^{q+1} > \lambda_- \simeq -1$.  Finally, the consensus time should scale as
$\tau \sim \frac{\ln N}{q(1-\omega)^{q+1}}$, as quoted in Eq.~(\ref{tau-q-g-1}).

\section{Continuum equation for $\langle \phi \rangle_f$} 
\label{continuum-eq}

In this section we derive an equation for the time evolution of the mean of a generic function $\phi(p)$ over the population of particles, expressed as  
\begin{eqnarray}
\langle \phi \rangle_f(t) \equiv \int_0^1 \phi(p) f(p,t) dp = \sum_{k=0}^S n_k(t) \, \phi(kh),
\label{phi-ave}
\end{eqnarray}
where $f(p,t)$ is the propensity distribution at time $t$.  Taking the time derivative of Eq.~(\ref{phi-ave}) gives
\begin{eqnarray} 
\label{dphi-ave-dt}
\frac{d}{dt} \langle \phi \rangle_f  &=& \sum_{k=0}^S \frac{d n_k}{dt}  \phi(kh) \nonumber \\ 
&=& n_0'(t)\phi(0)+n_S'(t)\phi(1) - \sum_{k=1}^{S-1}n_k\phi(kh) \nonumber \\ 
&& +\sum_{k=1}^{S-1} n_{k-1}\phi(kh)\sum_j n_jP_{k-1,j}
+\sum_{k=1}^{S-1} n_{k+1}\phi(kh)(1-\sum_j n_jP_{k+1,j}). 
\end{eqnarray} 
We then write 
\begin{eqnarray*}
	\phi(kh) &=& \phi((k-1)h) + h\phi'((k-1)h) + \frac{h^2}{2}\phi''((k-1)h) + O(h^3)  ~~~\mbox{and} \\
	\phi(kh) &=& \phi((k+1)h) - h\phi'((k+1)h) + \frac{h^2}{2}\phi''((k+1)h) + O(h^3), 
\end{eqnarray*} 
and replace in the 1st and 2nd summations of Eq.~(\ref{dphi-ave-dt}) respectively. 
After a change of indices and up to $o(h^2)$ terms we obtain 
\begin{eqnarray*}
	&& \sum_{k=1}^{S-1} n_{k-1}\phi(kh)\sum_j n_jP_{k-1,j}+\sum_{k=1}^{S-1} n_{k+1}\phi(kh)(1-\sum_j n_jP_{k+1,j})
	- \sum_{k=1}^{S-1}n_k\phi(kh) \\ 
	&& = \sum_{k=0}^{S-2} n_k(\phi(kh)+h\phi'(kh)+\frac12 h^2\phi''(kh))\sum_j n_jP_{k,j}
	+ \sum_{k=2}^S n_k(\phi(kh)-h\phi'(kh)+\frac12 h^2\phi''(kh)) (1-\sum_j n_jP_{kj}) 
	- \sum_{k=1}^{S-1}n_k\phi(kh) \\ 
	&& =: A+B
\end{eqnarray*}
with 
\begin{eqnarray*}
	A & = & 
	\sum_{k=0}^{S} n_k(\phi(kh)+h\phi'(kh)+\frac12 h^2\phi''(kh))\sum_j n_jP_{k,j}
	+ \sum_{k=0}^S n_k(\phi(kh)-h\phi'(kh)+\frac12 h^2\phi''(kh)) (1-\sum_j n_jP_{kj}) 
	-  \sum_{k=0}^S n_k\phi(kh) 
\end{eqnarray*}
and 
\begin{eqnarray*}
	B & = & 
	- n_{S-1}\Big( \phi(1-h)+h\phi'(1-h)+\frac12 h^2\phi''(1-h) \Big) \sum_j n_j P_{S-1,j} \\ 
	&& - n_S \Big( \phi(1)+h\phi'(1)+\frac12 h^2\phi''(1) \Big) \sum_j n_j P_{S,j} + n_S \phi(1) \\ 
	&& - n_0\Big( \phi(0)-h\phi'(0)+\frac12 h^2\phi''(0) \Big)  (1- \sum_j n_j P_{0,j}) + n_0\phi(0) \\ 
	&& - n_1\Big( \phi(h)-h\phi'(h)+\frac12 h^2\phi''(h) \Big) (1- \sum_j n_j P_{1,j}). 
\end{eqnarray*}
Notice that $A$ is equal to 
\begin{eqnarray}\label{A}
A & = & 
h\sum_{k=0}^S n_k \phi'(kh)\Big(2\sum_j n_jP_{k,j}-1\Big) 
+ \frac12 h^2 \sum_{k=0}^S n_k\phi''(kh) \nonumber \\ 
& = & h\int \phi'(p)v(p)\,f(p) \, dp + \frac12 h^2 \int \phi''(p)\,f(p) \, dp.
\end{eqnarray}
Writing that 
$$ \phi(1-h)+h\phi'(1-h)+\frac12 h^2\phi''(1-h) = \phi(1)+o(h^2),  $$ 
$$ \phi(h)-h\phi'(h)+\frac12 h^2\phi''(h) = \phi(0)+o(h^2)  $$ 
we obtain 
\begin{equation}\label{B}
B +n_0'(t)\phi(0)+n_S'(t)\phi(1)
= hn_0\Big(\phi'(0)-\frac12 h \phi''(0)\Big)(1-\sum_j n_jP_{0,j})
- hn_S\Big(\phi'(1)+\frac12 h \phi''(1)\Big)\sum_j n_jP_{S,j} 
\end{equation}
Replacing \eqref{A} and \eqref{B} in Eq.~(\ref{dphi-ave-dt})  leads to Eq.~(\ref{dphidt}) quoted in the main text.

\section{Linear stability analysis in the continuum limit $S \to \infty$}
\label{stability-delta-proof}

According to the discussion in section~\ref{stability-delta}, the asymptotic behavior of $f(p,t)$ results from the competition between the transport of $\gamma(p,t)$ in $(0,1)$ by the vector-field $V[\delta(p)]$ that drives the particles to the borders ($p=0$ and $p=1$) and the reentering of particles from the borders back to the $(0,1)$ interval.  Since 
\begin{eqnarray}
	V[\delta(p)](p)=2\mathbf{P}^+(p,0)-1 = \frac{2(\omega p)^q}{(\omega p)^q+(1-\omega p)^q}-1, 
\end{eqnarray}	
we have that for $q>0$ is 
\begin{equation}\label{Equ12}
V[\delta](p) < 0 \qquad \Longleftrightarrow \qquad \omega p<\frac12, 
\end{equation}
whereas for $q<0$ is 
\begin{equation}\label{Equ12}
V[\delta](p) < 0 \qquad \Longleftrightarrow \qquad \omega p>\frac12.  
\end{equation}
Thus when $q<0$ the vector-field $V[\delta](p)$ is positive for any small $p>0$, driving the particles to the right and thus $f(p,t) \not \to \delta(p)$.  We can then conclude that $\delta(p)$ is unstable for the linearized Eq.~(\ref{Equ11}) when $q<0$. 

Let us now assume that $q>0$ and that $\gamma(p,0)$ is supported in $[0,\Delta]$ with 
$\Delta <  \min\{(2\omega)^{-1},1\}$.  Then $ V[\delta](p)<0$ for any $p$ in the support of $\gamma(p,0)$.  It follows that $\gamma(p,t)$ is supported in $[0,\Delta]$ for any $t \ge 0$, in particular 
$\gamma(p=1,t)=0$.   Since $\varepsilon(t)\in [0,1]$ and  $\gamma(p,t)$ is a probability measure, 
there exist $\varepsilon_\infty:=\lim_{t\to +\infty}\varepsilon(t)$ 
and $\gamma_\infty:=\lim_{t\to +\infty} \gamma(p,t)$ (at least for a subsequence of times going to $+\infty$).  Hence $f(p,t)\to f_\infty:=(1-\varepsilon_\infty)\delta(p) + \varepsilon_\infty\gamma_\infty$. 
We want to prove that $f_\infty=\delta(p)$, i .e, that $\varepsilon_\infty=0$ or that $\gamma_\infty=\delta(p)$.  Suppose by contradiction that $\varepsilon_\infty>0$ and $\gamma_\infty\neq \delta(p)$.  Then the rhs of Eq.~(\ref{Equ11}) with $\varepsilon_\infty$ and $\gamma_\infty$ in place of $\varepsilon(t)$ and $\gamma(p,t)$ respectively
must be zero:  
\begin{equation}\label{Equ12}
\Big\langle V[\delta](p) \, \phi'(p) \Big \rangle_{\gamma_\infty} + \phi'(0) B_0[\gamma_\infty]  = 0 
\qquad \text{for any $\phi$. }
\end{equation}
Any $p$ in the support of $\gamma_\infty$ satisfies $0\le p\le \Delta$.  Taking $\Delta$ small, we can approximate
\begin{eqnarray*}
	&& \phi'(p)=\phi'(0)+o(1), \\ 
	&& \mathbf{P}^+(p,0) \simeq (\omega p)^q = o(1), \\ 
	&& \mathbf{P}^+(0,p) \simeq (1-\omega)^q p^q = o(1), 
\end{eqnarray*}
where $o(1)\to 0$ as $\Delta\to 0$. It follows that 
\begin{eqnarray*}
	&& V[\delta](p) = 2 \mathbf{P}^+(p,0) -1 = -1+o(1), \\
	&& B_0[\gamma_\infty] = \int_0^\Delta \mathbf{P}^+(0,p') \, \gamma_\infty(p') \, dp'= o(1).
\end{eqnarray*}
Hence \eqref{Equ12} becomes $ -\phi'(0) + o(1) = 0$ for any $\phi$.  Taking $\phi$ such that $\phi'(0)=1$, we obtain a contradiction for $\Delta$ small enough.  Thus any limit of $f(p,t)$ as $t \to \infty$ must be 
$\delta(p)$.  Since the set of probability measures on $[0,1]$ is compact, we conclude that $f(p,t) \to \delta(p)$. Therefore, $\delta(p)$ is locally asymptotically stable for $q>0$ and unstable for $q<0$, for the 
the linearized Eq.~(\ref{Equ11}), as mentioned in section~\ref{stability-delta}.

\bibliography{references}

\end{document}